\documentclass[%
 reprint,
 superscriptaddress,
 showpacs,
 aps,
 prstab,
]{revtex4-1}

\usepackage[english]{babel}
\usepackage{amsmath,amssymb,amstext,amsfonts}
\RequirePackage{graphicx}
\usepackage{upgreek}

\RequirePackage{titlesec}

\RequirePackage{enumitem}
\setitemize{noitemsep,topsep=0ex,parsep=\parskip,partopsep=0pt,leftmargin=*}
\setenumerate{topsep=0pt,parsep=\parskip,partopsep=0pt,leftmargin=*,font=\bfseries\sffamily}
\setlist[description]{topsep=0pt,parsep=\parskip,partopsep=0pt,leftmargin=0pt,labelindent=0pt,font=\bfseries\small}

\usepackage{hyperref}
\hypersetup{
    colorlinks=true,
    linkcolor=blue,
    filecolor=blue,      
    urlcolor=blue,
    citecolor=blue,
}
\usepackage[mathlines,left]{lineno}
\graphicspath{{./figs/}}
\DeclareGraphicsExtensions{.pdf,.png}

\begin{document}

\title{Hybrid LWFA $\vert$ PWFA staging as a beam energy and brightness transformer: conceptual design and simulations}

\author{A. Martinez de la Ossa}
\email[]{alberto.martinez.de.la.ossa@desy.de}
\affiliation{Deutsches Elektronen-Synchrotron DESY, 22607 Hamburg, Germany}%
\author{R. W. A\ss mann}
\affiliation{Deutsches Elektronen-Synchrotron DESY, 22607 Hamburg, Germany}%
\author{M. Bussmann}
\affiliation{Helmholtz-Zentrum Dresden-Rossendorf HZDR, 01328 Dresden, Germany}%
\author{S. Corde}
\affiliation{LOA, ENSTA ParisTech - CNRS - \'{E}cole Polytechnique - Universit\'{e} Paris-Saclay, France}%
\author{J. P. Couperus Cabada\u{g}}
\affiliation{Helmholtz-Zentrum Dresden-Rossendorf HZDR, 01328 Dresden, Germany}%
\author{A. Debus}
\affiliation{Helmholtz-Zentrum Dresden-Rossendorf HZDR, 01328 Dresden, Germany}%
\author{A. D\"{o}pp}
\affiliation{Ludwig-Maximilians-Universit\"{a}t M\"{u}nchen, Am Coulombwall 1, 85748 Garching, Germany}%
\author{A. Ferran Pousa}
\affiliation{Deutsches Elektronen-Synchrotron DESY, 22607 Hamburg, Germany}%
\author{M. F. Gilljohann}
\affiliation{Ludwig-Maximilians-Universit\"{a}t M\"{u}nchen, Am Coulombwall 1, 85748 Garching, Germany}%
\author{T. Heinemann}
\affiliation{Deutsches Elektronen-Synchrotron DESY, 22607 Hamburg, Germany}%
\affiliation{Scottish Universities Physics Alliance, Department of Physics, University of Strathclyde, Glasgow G4 0NG, UK}
\author{B. Hidding}
\affiliation{Scottish Universities Physics Alliance, Department of Physics, University of Strathclyde, Glasgow G4 0NG, UK}
\author{A. Irman}
\affiliation{Helmholtz-Zentrum Dresden-Rossendorf HZDR, 01328 Dresden, Germany}%
\author{S. Karsch}
\affiliation{Ludwig-Maximilians-Universit\"{a}t M\"{u}nchen, Am Coulombwall 1, 85748 Garching, Germany}%
\author{O. Kononenko}
\affiliation{LOA, ENSTA ParisTech - CNRS - \'{E}cole Polytechnique - Universit\'{e} Paris-Saclay, France}%
\author{T. Kurz}
\affiliation{Helmholtz-Zentrum Dresden-Rossendorf HZDR, 01328 Dresden, Germany}%
\author{J. Osterhoff}
\affiliation{Deutsches Elektronen-Synchrotron DESY, 22607 Hamburg, Germany}%
\author{R. Pausch}
\affiliation{Helmholtz-Zentrum Dresden-Rossendorf HZDR, 01328 Dresden, Germany}%
\author{S. Sch\"{o}bel}
\affiliation{Helmholtz-Zentrum Dresden-Rossendorf HZDR, 01328 Dresden, Germany}%
\author{U. Schramm}
\affiliation{Helmholtz-Zentrum Dresden-Rossendorf HZDR, 01328 Dresden, Germany}%

\begin{abstract}
  We present a conceptual design for a hybrid laser-to-beam-driven plasma wakefield accelerator.
  In this setup, the output beams from a laser-driven plasma wakefield accelerator (LWFA) stage
  are used as input beams of a new beam-driven plasma accelerator (PWFA) stage.
  In the PWFA stage a new witness beam of largely increased quality can be produced
  and accelerated to higher energies.
  The feasibility and the potential of this concept is shown
  through exemplary particle-in-cell simulations.
  In addition, preliminary simulation results for a proof-of-concept
  experiment at HZDR (Germany) are shown.
\end{abstract}



\maketitle

\section{Introduction}
Plasma-based accelerators driven by either an intense laser pulse~\cite{Tajima1979} (LWFA)
or a highly relativistic charged particle beam~\cite{Veksler1956,Chen1985,Chen1987} (PWFA),
generate and sustain accelerating fields orders of magnitude
higher than those achievable with conventional radiofrequency technology,
offering a path towards highly-compact and cost-effective particle accelerators,
with multiple applications in science, industry and medicine.

In both LWFAs and PWFAs either a laser or a particle beam
excites wakefields in an initially neutral and homogeneous plasma.
These wakefields propagate at the velocity of the driver and oscillate
at the plasma frequency, $\omega_p = \sqrt{n_{p} e^2/m\epsilon_{0}}$,
where $e$ is the elementary charge, $m$ is the mass of the electron,
$\epsilon_{0}$ is the vacuum permittivity and $n_{p}$ is the unperturbed
plasma electron density.
Sufficiently intense drivers generate the wakefields in the blowout regime,
completely expelling all plasma electrons from their propagation path,
and forming a clear ion cavity with a length given approximately by 
$\lambda_p = 2\pi/k_p \simeq 33~\mathrm{\upmu m}\times n_p [10^{18}\mathrm{cm^{-3}}]^{-1/2}$, 
with $k_p = \omega_p/c$, the wakefield oscillation wavenumber, and $c$ the speed of light.
Inside this cavity accelerating fields exceeding 
$E_0 = (mc/e)\,\omega_p \simeq 96~(\mathrm{GV/m})\times n_p [10^{18}\mathrm{cm^{-3}}]^{1/2}$ 
are generated, as well as an uniform focusing gradient 
$K = (m/2ec)\,\omega_p^2 \simeq 30~(\mathrm{MT/m}) \times n_p [10^{18}\mathrm{cm^{-3}}]$,
enabling ideal conditions for the acceleration and transport of electron beams
to $\mathrm{GeVs}$ energies within $\mathrm{cm}$-scale distances~\cite{Rosenzweig1991,Lotov2004,Lu2006}.

Although LWFAs and PWFAs share the same working principle,
they possess some fundamental differences inherited from
the completely different nature of their drivers:
in LWFAs, a drive laser with frequency $\omega_0$ propagates with
a reduced group velocity in the plasma, $v_0 = c\, \sqrt{1-(\omega_p/\omega_0)^2}$, 
which results in a progressive dephasing between the electrons being accelerated
at near the speed of light and the wakefields.
Besides, although the laser propagation can be self-\cite{Thomas2007,Lu2007}
or externally guided~\cite{Spence2000,Shalloo2018},
the etching and diffraction effects on the laser pulse substantially affect
the wakefield excitation conditions~\cite{Streeter2018}, 
which together with the dephasing effect, harden the required control
over the witness acceleration process for applications demanding witness beams
with low energy spread and emittance. 
By contrast, PWFAs driven by highly relativistic electron beams 
are not affected by dephasing (the drive beam propagates at nearly the speed of light),
and when in the blowout regime, the wakefield excitation state remains essentially
unaltered over most of the propagation, making possible to find optimal conditions
of beam loading for a minimal correlated energy spread~\cite{Lotov2005,Tzoufras2008},
which hold over the entire acceleration process.
Moreover, operating PWFAs in the blowout regime enables novel injection techniques
specifically designed to deliver ultra-low emittance witness beams
~\cite{Hidding2012,Li2013,MartinezdelaOssa2013,MartinezdelaOssa2015,Wittig2015,MartinezdelaOssa2017,Manahan2017}, 
that can be efficiently accelerated within the blowout plasma wake, 
therefore allowing for a substantial improvement of the energy and the quality
of the generated witness beams with respect to the drivers.

Other significant differences between LWFAs and PWFAs arise from their practical implementation:
while LWFAs are operated by relatively compact high-power laser systems,
PWFAs typically require costly km-scale electron linacs~\cite{Hogan2010,Aschikhin2016}.
Thanks to the proliferation of multiple laser laboratories around the world capable of producing
short laser pulses with hundreds $\mathrm{TW}$ of peak power~\cite{Strickland1985}, 
LWFAs have been greatly advanced over the last decades.
Thus, important milestones, such as the realization of quasi-monoenergetic
electron spectra~\cite{Mangles2004,Geddes2004,Faure2004}, 
GeV-class beams~\cite{Leemans2006,Wang2013,Leemans2014},
enhanced stability~\cite{Osterhoff2008,Hafz2008}, 
controlled injection techniques for tunability~\cite{Faure2006,Rowlands-Rees2008,Pak2010,Schmid2010,Gonsalves2011,Buck2013,Mirzaie2015}, 
and the application of the generated beams to drive compact synchrotron~\cite{Schlenvoigt2008}, XUV~\cite{Fuchs2009}
and X-ray sources~\cite{Kneip2010,Khrennikov2015,Dopp2017},
have validated plasma-based acceleration as a promising technique for future accelerators.
By contrast, only few linac facilities are nowadays capable and ready to perform PWFAs,
and in spite of the important advances achieved
at SLAC~\cite{Blumenfeld2007,Oz2007,Litos2014,Corde2015,Corde2016},
DESY~\cite{Gross2018,Loisch2018} and CERN~\cite{Adli2018},
PWFA research is handicapped by the comparably limited number of facilities which allow to realize PWFA.

In the context of this article, LWFAs are thought to provide a relatively compact and
affordable source of high-energy electron beams for driving PWFAs,
in order to allow an easier and widespread access to PWFA technology for research and applications.
Besides, the LWFA-produced beams have some unique properties as PWFA drivers.
LWFAs are now proven to routinely produce GeV-class and highly compressed beams~\cite{Lundh2011,Heigoldt2015,Couperus2017,Irman2018,Li2017}, 
with peak currents well above the minimal requirements for enabling internal injection techniques
and efficient acceleration in PWFAs.
By contrast, achieving the drive beam energy and current required to enable a strong blowout regime in PWFAs
imposes challenging operating conditions for linacs upon compressing and transporting the beam.
Moreover, although the state-of-the-art emittance and energy spread of the LWFA-generated electron
beams are as of yet insufficient for applications demanding a high beam quality (e.g. FELs),
these characteristics do not inhibit their usage as driver beams for PWFAs and may even offer improved stability~\cite{Mehrling2017,MartinezdelaOssa2018}.

\begin{figure*}[t]
\centering\includegraphics[width=1.0\textwidth]{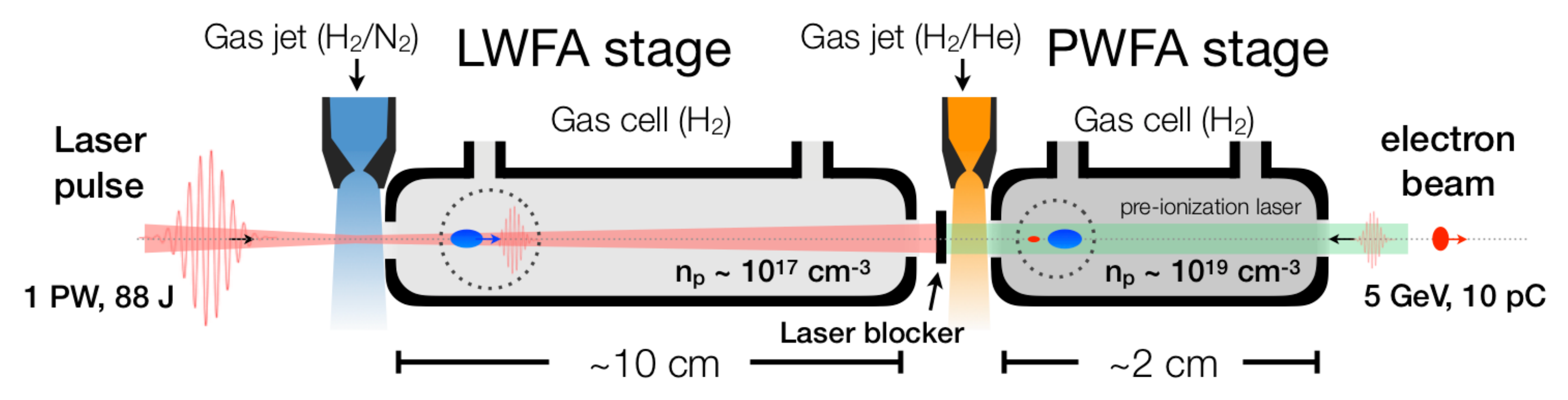}
\caption{Schematic of a hybrid LWFA $\vert$ PWFA staged setup with ionization injection.
  The setup consists of two quasi-identical plasma acceleration stages coupled to each other.
  In each stage, the injection of a witness beam is induced by field-ionization from the dopant
  species contained in the gas jet. Immediately after the gas jet, a longer plasma cell
  with no dopant is used to further accelerate the generated witness beam.
}
\label{fig:scheme}
\end{figure*}
Using the output of a LWFA to perform a PWFA was already proposed
as a beam-energy afterburner, where two distinct electron bunches produced in an LWFA 
were thought to be used as a driver/witness pair in a PWFA stage~\cite{Hidding2010}.
Also a mode transition from LWFA to PWFA in one single stage has been proposed as a way
of boosting the energy~\cite{Pae2010,Masson-Laborde2014} and/or the betatron radiation~\cite{Corde2011,Dong2018,Ferri2018}
of the LWFA-produced electron beams.
The concept developed here is fundamentally different, as the witness beam is  
generated in the PWFA stage in a controlled manner,
and the hybrid LWFA $|$ PWFA staging is conceived as a beam energy and brightness transformer. 
This possibility of using LWFA-produced electron beams to drive a PWFA
for the generation of superior quality beams has been also indicated
in \cite{Hidding2012,MartinezdelaOssa2015,Manahan2017}.  
So far, important experimental milestones such that LWFA beams can generate
wakefields in a subsequent plasma stage have been
demonstrated~\cite{Kuschel2016,Chou2016,Gilljohann2018},
but no acceleration of a distinct witness beam in an LWFA-driven PWFA has ever
been reported.

In this work, we further study the potential of an LWFA-driven PWFA (LPWFA) 
by means of idealized setups simulated with the particle-in-cell (PIC) code OSIRIS~\cite{Fonseca2002,Fonseca2008,Fonseca2013}.
Specifically, we do not aim towards full predictive start-to-end simulations here,
but for testing the feasibility of such scheme in the light of current experimental capabilities.
In the scheme considered, the output beam from an LWFA stage is used as the driver of a PWFA stage,
where a new witness beam of superior quality is generated and accelerated to higher energies.
This scenario benefits from the advantages unique to each method, 
particularly exploiting the capability of PWFA schemes to provide energy-boosted and 
high-brightness witness beams, while the LWFA stage fulfills the demand for
a compact source of relativistic high-current electron bunches required as PWFA drivers.
In essence, the PWFA stage operates as a beam brightness and energy transformer of
the LWFA output, aiming to reach the demanding beam quality requirements
of accelerator-driven light sources~\cite{Huang2012,Maier2012,Hidding2014,Dimitri2015},
without sacrificing the small spatial footprint and the relatively low cost offered by LWFAs.

\section{Conceptual design and simulations}
\label{sec:concept}
We start describing a conceptual design for an LPWFA consisting of two
quasi-identical plasma acceleration modules coupled to each other with a minimal distance in between
(Figure~\ref{fig:scheme}).
Each plasma module contains a gas jet (supersonic gas nozzle) at the front,
which is fed with a low-ionization-threshold (LIT) gas species (e.g. hydrogen),
doped with a high-ionization-threshold (HIT) gas species at variable concentration.
In each stage, the system triggers the injection of a witness beam
via field-ionization from the dopant HIT species contained in the gas jet.
The injection stops once the gas jet column is over.
Immediately after the gas jet,
a longer plasma cell composed only of the LIT gas species 
is used to further accelerate the generated witness beam.

The first plasma stage is driven by a high-power laser (LWFA),
optimized for the production of a highly relativistic (order $\mathrm{GeV}$ energy)
and high-current $\gtrsim 10~\mathrm{kA}$ electron beam via ionization injection~\cite{Rowlands-Rees2008,Pak2010}.
The LWFA electron beam is then used as driver of the subsequent plasma stage (PWFA),
where a new high-quality witness beam is produced via wakefield-induced ionization injection~\cite{MartinezdelaOssa2013,MartinezdelaOssa2015}
and then boosted to high-energies.
Both plasma stages need to be sufficiently close to each other,
such that the LWFA-beam can be refocused into the second plasma by means of its self-driven plasma wakefields.
A thin slab of solid material (e.g. aluminum, steel, kapton, etc.) is placed in the beginning of the second jet
in order to remove the laser from the second stage, while letting the electron beam pass through.
In addition, a counter-propagating low-intensity laser can be used to 
preionize the LIT gas species in the PWFA stage, 
in order to facilitate the beam focusing and enhance the blowout formation.
Typically, the plasma density in the second stage needs to be increased substantially with respect to
the one in the first stage, as it needs to be matched to the short length of the LWFA beam
for an efficient injection and acceleration process.
By operating the PWFA stage with a short and high-current electron beam at these high plasma densities, 
it is possible to generate ultra-short electron beams (sub-femtosecond duration),
which double the initial energy of the LWFA beam in just few cm of acceleration,
for the parameters considered.
Moreover, the emittance of the newly generated PWFA beam can be on the order of hundred nanometers,
at the same time that its current can reach values of tens of kiloamps, as required for optimum beam loading.
The brightness enhancement of the PWFA beam with respect to the LWFA beam can be about five orders of magnitude.

\subsection{First stage: LWFA with ionization injection}
\label{sec:lwfa}
A first PIC simulation employing the code OSIRIS~\cite{Fonseca2002,Fonseca2008,Fonseca2013}
was performed for the LWFA stage.
For the drive laser, we considered a Ti:Sa system ($\lambda_0 = 800~\mathrm{nm}$),
with a peak power of $P_0 = 98 \mathrm{TW}$.
The parameters of the laser in the simulation resemble those of the DRACO laser system~\cite{Schramm2017}
at HZDR (Helmholtz-Zentrum Dresden-Rossendorf),
which was successfully utilized to perform LWFA experiments with ionization-injection~\cite{Couperus2017,Irman2018}.
The laser intensity envelope is Gaussian in both longitudinal and transverse directions.
The pulse duration (fwhm) is $\tau_0 = 27~\mathrm{fs}$, while the spot size at focus (waist) is $w_0 = 17\,\mathrm{\upmu m}$,
yielding a peak normalized vector potential of $a_0 = 3.18$,
which correspond to a peak intensity of $I_0 = 2.16\times 10^{19}~\mathrm{W/cm^2}$.
The total energy in the laser pulse amounts to $2.8~\mathrm{J}$.
The laser is linearly polarized in the $x$ direction.
The plasma density was chosen to be $n_p = 2 \times 10^{18}~\mathrm{cm^{-3}}$ in order
to provide near-resonance wakefield excitation
($k_p\sigma_{z,0} \approx 1$, with $\sigma_{z,0} \simeq c\tau_0/2.35$, the rms length of the laser pulse)
and relativistic self-guiding ($k_pw_0 \approx 2\sqrt{a_0}$) of the laser intensity envelope~\cite{Lu2007}.
The plasma profile starts with a short Gaussian up ramp (shorter than the plasma wavelength) from vacuum to the plateau density at $n_p$.
Then the plasma density continues uniform at $n_p$ for the rest of the simulation.
The first part of the plasma profile is doped at $1\%$ with nitrogen (N), up to $200~\mathrm{\upmu m}$ from the start of the plateau.

\begin{figure*}[!t]
\centering\includegraphics[width=1.0\textwidth]{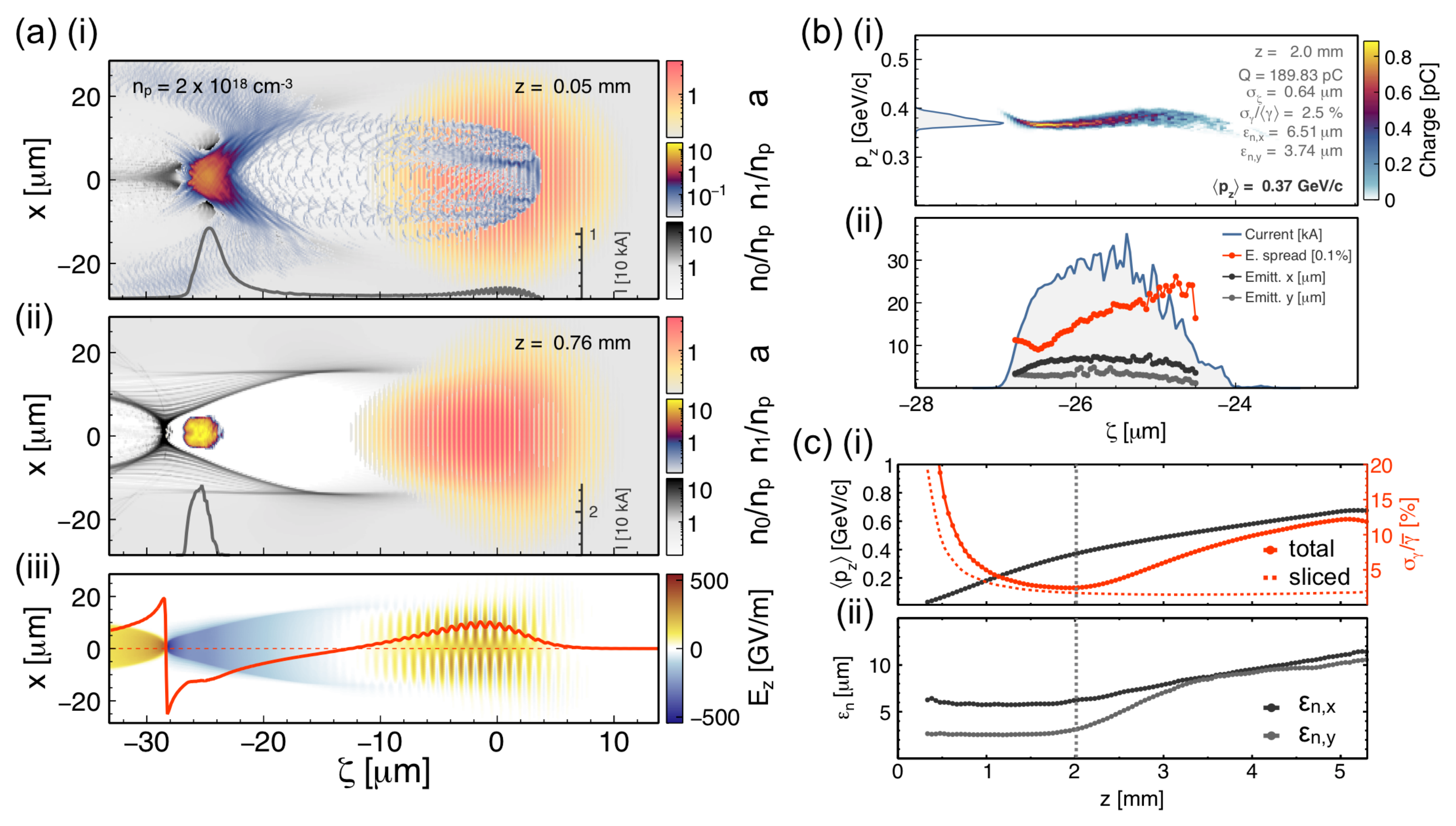}
\caption{3D OSIRIS simulation for an LWFA stage with ionization injection.
  (a)~(top): electron density on the central $x-z$ plane of the simulation during the injection process,
  at $z=0.05~\mathrm{mm}$. (a)~(middle): same quantities once 
  the dopant section has been passed, at $z=0.76~\mathrm{mm}$.
  The electron densities of the plasma (gray) and the high levels of nitrogen (blue/yellow) are shown.
  Also the magnitude of the normalized vector potential of the laser is shown (orange/red).
  The dark gray line at the bottom shows the charge per unit of length of the injected electrons.
  (a)~(bottom): longitudinal electric field, $E_z$, also at $z=0.76~\mathrm{mm}$.
  The red outline represents the on-axis values.
  (b)~(top): longitudinal phase space of the witness beam,
  and (b)~(bottom): sliced values of the current (blue),
  relative energy spread (red), and normalized emittance in the $x$ (dark grey), and $y$ (light grey) planes.
  (c)~(top): evolution of the average longitudinal momentum of the beam (dark grey),
  the total relative energy spread (red),
  and the average sliced relative energy spread (dashed red), with the propagation distance.
  (c)~(bottom): evolution of the projected normalized emittance of the beam in the $x$ (dark grey), and $y$ (light grey) planes.
}
\label{fig:lwfasim}
\end{figure*}
Figure~\ref{fig:lwfasim}(a)~(top) shows a snapshot of the simulation when the drive laser is traversing the doped section.
The electron densities of the plasma and the high levels ($\mathrm{N}^{5+}$ and $\mathrm{N}^{6+}$) of nitrogen are shown.
The low levels of nitrogen were not included in the simulation as they play a negligible role in the injection process 
and including them substantially increases the computational cost.
Figure~\ref{fig:lwfasim}(a)~(middle) shows another snapshot after the dopant section has been passed,
at a propagation distance of $z = 0.76~\mathrm{mm}$ after the start of the density plateau.
A witness beam has been injected, composed by electrons from the high levels of nitrogen.
The current profile of this beam is also shown as a dark gray line.
This beam features a peak current of $35~\mathrm{kA}$, $190~\mathrm{pC}$ charge and $6~\mathrm{fs}$ duration (fwhm).
Figure~\ref{fig:lwfasim}(a)~(bottom) shows the longitudinal electron field also at $z = 0.76~\mathrm{mm}$.
The accelerating field over the witness bunch at this point is around $200~\mathrm{GV/m}$.

Figure~\ref{fig:lwfasim}(b) shows in details the phase space of the witness bunch after
$2.0~\mathrm{mm}$ of propagation.
The average energy of the beam is $\bar{\gamma}_\mathrm{w} = 370~\mathrm{MeV}/\mathrm{mc^2}$,
and therefore, the average accelerating field up to this point is around $185~\mathrm{GV/m}$.
The total energy spread is $2.5\%$, while the sliced energy spread is around $2\%$.
Further acceleration of the witness beam is still possible since the system did not reach neither the dephasing
nor the depletion length, the two main limiting factors to the energy gain of the witness beam in an LWFA~\cite{Lu2007}.
According to the scalings for short and intense drive lasers operating in the non-linear regime
under self-guiding conditions~\cite{Lu2007},
the acceleration distance is limited to the pump-depletion length in this case:
$L_{dp} \approx (\omega_0/\omega_p)^{2}\,(\omega_p\tau_0)\,(c/\omega_p) \simeq 7~\mathrm{mm}$,
while the average accelerating field, accounting for dephasing, is $\bar{E}_z^\mathrm{w} \approx E_0\,\sqrt{a_0}/2 \simeq 120~\mathrm{GV/m}$.
Thus, the expected total energy gain after the pump-depletion length is approximately given by
$\bar{\gamma}_\mathrm{w} \approx L_{dp}\,\bar{E}_z^\mathrm{w} = (\omega_0/\omega_p)^{2}\,(\omega_p\tau_0)\,\sqrt{a_0}/2 \simeq 0.84~\mathrm{GeV}/\mathrm{mc^2}$.
We note that this expression for $\bar{E}_z^\mathrm{w}$ holds for a beam-unloaded case. 
However, the simulation shown in here is optimized for the production of high-current witness beam,
and thus, there is a substantial beam-loading effect, which overall diminishes the high slope of the
accelerating gradient along the witness beam.
As a consequence, the maximum energy gain in the witness beam is observed after only $5~\mathrm{mm}$ of propagation,
reaching values of $\sim 0.7~\mathrm{GeV}$.

\label{sec:scaling}
\begin{table*}[!t]
  \begin{tabular}{llll}
    Plasma density         &   $n_p$   & $2 \times 10^{18}\mathrm{cm}^{-3}$ & $2 \times 10^{17}\mathrm{cm}^{-3}$  \\
    \hline\hline
    Laser peak intensity   & $a_0 \propto 1$           & $3.18$ &  $3.18$\\
    Laser spot size (norm.) & $w_0 \propto n_p^{-1/2}$    & $17\,\mathrm{\upmu m}~(4.5)$ & $54~\mathrm{\upmu m}~(4.5)$ \\
    Laser pulse duration (norm.) & $\tau_0\propto n_p^{-1/2}$    & $27~\mathrm{fs}~(2.15)$ & $85~\mathrm{fs}~(2.15)$ \\
    Laser power            & $P_0 \propto n_p^{-1}$   & $98~\mathrm{TW}$ & $980~\mathrm{TW}$ \\
    Laser energy           & $E_0 \propto n_p^{-3/2}$ & $2.8~\mathrm{J}$ & $88~\mathrm{J}$ \\
    \hline
    Acceleration distance  & $L_\mathrm{acc} \propto n_p^{-3/2}$ & $2.0~\mathrm{mm}$ & $6.3~\mathrm{cm}$ \\
    Accelerating field     & $E_z^\mathrm{w} \propto n_p^{1/2}$ & $185~\mathrm{GV/m}$ & $58~\mathrm{GV/m}$ \\
    Beam energy (mean)     & $\bar{\gamma}_\mathrm{w}\mathrm{mc^2} \propto n_p^{-1}$ & $370~\mathrm{MeV}$ & $3.7~\mathrm{GeV}$ \\
    Beam charge            & $Q_\mathrm{w}\propto n_p^{-1/2}$ & $190~\mathrm{pC}$  & $601~\mathrm{pC}$ \\ 
    Beam current           & $I_\mathrm{w}\propto 1$ & $30~\mathrm{kA}$  & $30~\mathrm{kA}$ \\ 
    Beam duration (fwhm)   & $\tau_\mathrm{w}\propto n_p^{-1/2}$ &  $6~\mathrm{fs}$  &  $19~\mathrm{fs}$ \\ 
    Beam norm. emittance   & $\epsilon_{n,\mathrm{w}}\propto n_p^{-1/2}$ &  $5~\mathrm{\upmu m}$  &  $16~\mathrm{\upmu m}$ \\
    Beam rel. energy spread  & $\sigma_\gamma^{\mathrm{w}}/\bar{\gamma}_{\mathrm{w}} \propto 1$ &  $2.5~\%$  &  $2.5~\%$ \\
    \hline
    \hline
  \end{tabular}
  \caption{Laser, plasma and beam parameters for the simulated case at $n_p = 2 \times 10^{18}\mathrm{cm}^{-3}$
    and the extrapolated case at $n_p = 2 \times 10^{17}\mathrm{cm}^{-3}$.}
  \label{tab:lwfasim}
\end{table*}
The time evolution of the witness beam parameters is shown in Figure~\ref{fig:lwfasim}~(c). 
The minimum relative energy spread is reached after $\sim 2~\mathrm{mm}$ of propagation.
At this point the plasma wakefields are well beam loaded by the witness beam,
and therefore, its total energy spread is relatively low ($2.5\%$) and comparable to the slice energy spread ($2\%$). 
However, due to the beam dephasing effect and the laser diffraction,
the optimal beam-loading conditions do not hold over the whole propagation,
leading to the increase of the correlated energy spread (energy chirp) as the beam is further accelerated.
The projected normalized emittance of the beam is also shown in Figure~\ref{fig:lwfasim}~(c).
Initially the emittance is approximately two times higher in the laser polarization plane,
reaching values of $\sim 6~\mathrm{\upmu m}$.
However, as the drive laser diffracts, the generated plasma wakefields diminish in intensity and
the blowout formation is not complete.
Due to the presence of plasma electrons in the first plasma oscillation bucket,
the witness beam alters the plasma currents by means of its space-charge fields,
generating in this way a longitudinal variation of the focusing,
which in turns causes an increase of the projected emittance in both transverse planes.

In summary, we show by means of a PIC simulation that electron beams
with tens of $\mathrm{kA}$ and hundreds of $\mathrm{MeV}$ can be produced in an LWFA with ionization injection,
when employing a $\sim100~\mathrm{TW}$ peak power laser system.
These results are in good agreement with recent experiments~\cite{Couperus2017,Irman2018,Li2017}.
In particular, in ref.~\cite{Couperus2017} it is shown that by controlling the amount of injected charge,
e.g. by changing the concentration of the dopant,
optimal beam-loading conditions could be found for the production of $220 \pm 40~\mathrm{pC}$ charge
electron beams, with $250 \pm 20~\mathrm{MeV}$ average energy and $15\%$ relative energy spread.

\subsubsection*{Scaling results to petawatt power lasers}
\label{sec:scaling}
In this section we show by means of well-known scaling rules~\cite{Lu2007} that by employing higher power lasers, 
these results can be extrapolated to the production of longer beams with higher charges,
and with average energies in the multi-GeV range.
In particular, we show an example for the EuPRAXIA design study~\cite{Walker2017}, 
which considers a drive laser pulse with $\sim 1~\mathrm{PW}$ peak power
operating a plasma at a 10 times smaller density, i.e. $n_p = 2 \times 10^{17}~\mathrm{cm}^{-3}$.
Performing a new PIC simulation for these parameters would require to increase the longitudinal resolution
by a factor $\sqrt{10}$,
which together with the corresponding increase in the time resolution (factor $\sqrt{10}$)
and the propagation distance (factor $10$),
would require about a factor $100$ more computing power, making this task extremely costly.
For this reason, we rely on scaling rules to estimate the results when the plasma density is varied,
while keeping $\omega_0$, $a_0$, $k_pw_0$ and $\omega_p\tau_0$ of the drive laser constant.
Since the sizes of the laser pulse are kept constant relative to the plasma wavelength, 
the laser peak power scales as $P_0 \propto n_p^{-1}$ for a constant $a_0$, 
while the total energy in the pulse, $E_\mathrm{laser} \simeq P_0\tau_0$, follows a $n_p^{-3/2}$ dependency.
On the other hand, the laser propagation distance in plasma goes as
$L_\mathrm{acc} \propto n_p^{-3/2}$, while the accelerating wakefield does as $\bar{E}_z \propto n_p^{1/2}$.
Therefore, the final energy of the witness beam scales as $\bar{\gamma}_\mathrm{w} \propto n_p^{-1}$.

\begin{figure*}[!t]
\centering\includegraphics[width=1.0\textwidth]{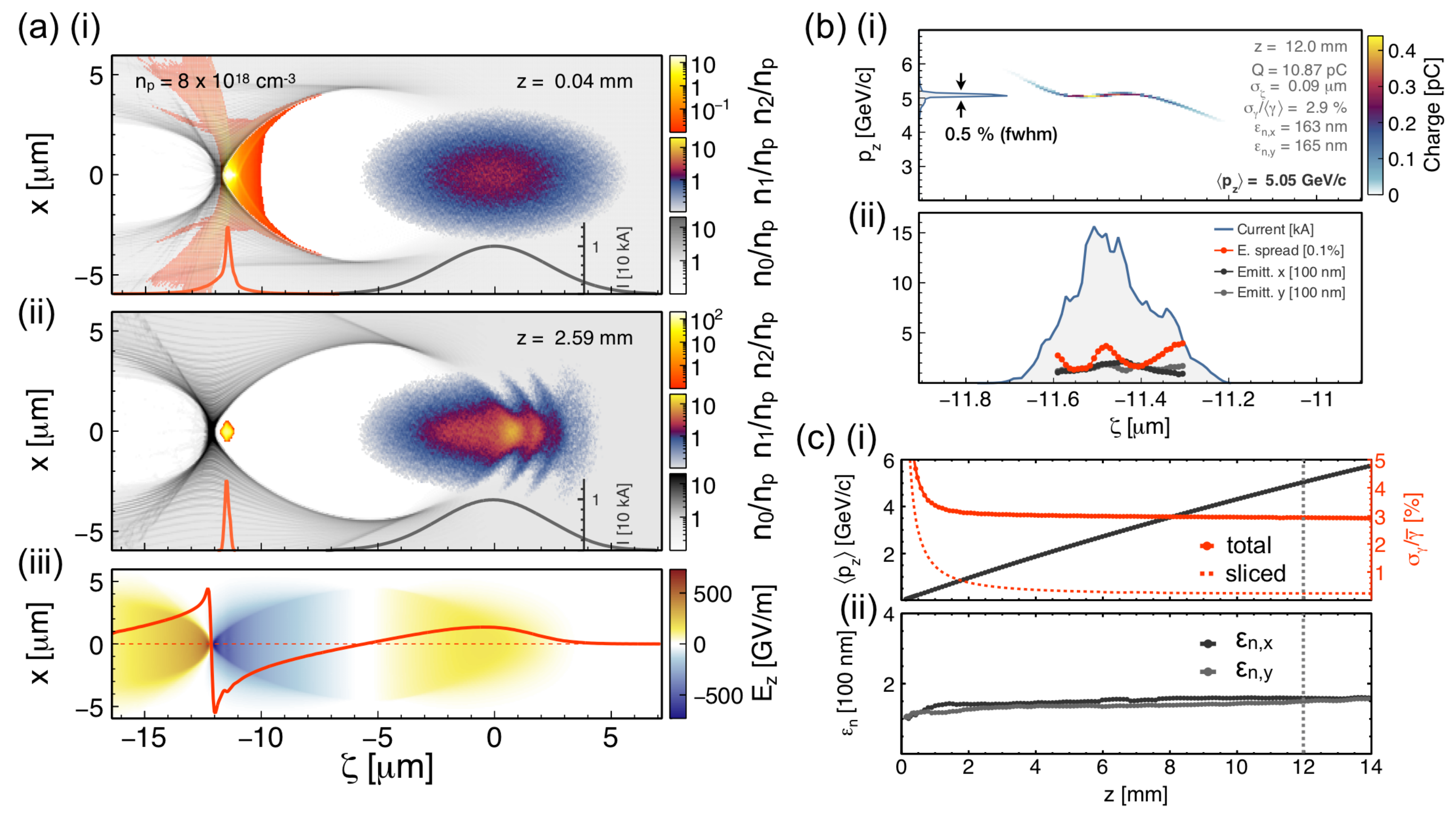}
\caption{3D OSIRIS simulation for a PWFA stage with wakefield-induced ionization injection.
  (a)~(top): electron density on the central $x-z$ plane of the simulation during the injection process,
  at $z=0.04~\mathrm{mm}$. (a)~(middle): same quantities once 
  the dopant section has been passed, at $z=2.6~\mathrm{mm}$.
  The electron densities of the plasma (gray) and the $\mathrm{He^+}$ level (orange/yellow) are shown.
  The dark gray and orange lines at the bottom show the charge per unit of length of the driver and
  the injected electrons, respectively.
  (a)~bottom: longitudinal electric field, $E_z$, also at $z=2.6~\mathrm{mm}$.
  The red outline represents the on-axis values.
  (b)~(top): longitudinal phase space of the witness beam after $12~\mathrm{mm}$ of propagation,
  and (b)~(bottom): sliced values of the current (blue),
  relative energy spread (red), and normalized emittance in the $x$ (dark grey) and $y$ (light grey) planes.
  (c)~(top): evolution of the average longitudinal momentum of the beam (dark grey),
  the total relative energy spread (red),
  and the average sliced relative energy spread (dashed red).
  (c)~(bottom): evolution of the projected normalized emittance of the beam in the $x$ (dark grey), and $y$ (light grey) planes.}
\label{fig:pwfasim}
\end{figure*}
Taking as a reference the witness beam generated in the 3D LWFA simulation described in Section~\ref{sec:lwfa},
we summarize in Table~\ref{tab:lwfasim} the expected beam parameters when the process is initiated by
a PW-class Ti:Sa laser in a plasma with $2 \times 10^{17}~\mathrm{cm}^{-3}$ density.
The corresponding laser pulse driving the process would have then the following parameters:
$P_0 = 980~\mathrm{TW}$, $\tau_0 = 85~\mathrm{fs}$ and $w_0 = 54~\mathrm{\upmu m}$.
$a_0$ is kept fixed and equals to $3.18$ for a Ti:Sa laser.
The total energy in the pulse would be then $88~\mathrm{J}$.
For the produced witness beam, we estimate an energy gain of $3.7~\mathrm{GeV}$
after a propagation distance of $6.3~\mathrm{cm}$.
This is the equivalent distance at which the witness beam had the lowest energy spread
in the simulated case (cf. Figure~\ref{fig:lwfasim}~(c)).
The current profile of the witness beam remains the same for the same degree of beam loading~\cite{Tzoufras2008}, 
but its length scales as $n_p^{-1/2}$, leading to a $19~\mathrm{fs}$ long (fwhm) beam.
Also the total charge, $Q_\mathrm{w}$, scales as the beam length, yielding $601~\mathrm{pC}$.
The normalized emittance of the witness beam, $\epsilon_{n,\mathrm{w}}$, is essentially defined by
the transverse extension of the initial distribution of the ionized electrons~\cite{Kirby2009,Schroeder2014},
and therefore is expected to also scale as $\epsilon_{n,\mathrm{w}} \propto n_p^{-1/2}$, 
yielding $16~\mathrm{\upmu m}$ for the extrapolated case.
Assuming that the total energy spread is dominated by the energy chirp
imprinted in the bunch due to a non-uniform accelerating field along the beam,
we expect this quantity to scale as the energy gain $\sigma_\gamma^{\mathrm{w}} \propto n_p^{-1}$,
and therefore, the relative energy spread $\sigma_\gamma^{\mathrm{w}}/\bar{\gamma}_\mathrm{w}$ would remain unchanged. 
We note that the total energy stored in the witness beam,
given by $Q_\mathrm{w} \bar{\gamma}_\mathrm{w} \propto n_p^{-3/2}$, 
scales as the laser pulse energy, and therefore, the laser-to-beam energy transfer efficiency remains constant.

\subsection{Second stage: PWFA with wakefield-induced ionization injection}
\label{sec:pwfa}
In this section we present an exemplary PIC simulation that shows that electron beams
with the same properties than those produced in LWFAs with ionization injection
can indeed be utilized as drivers of a PWFA.
In this LWFA-driven PWFA (LPWFA) stage, a new witness beam of largely superior quality
is injected by means of wakefield-induced ionization (WII) injection~\cite{MartinezdelaOssa2013,MartinezdelaOssa2015}
and then accelerated to substantially higher energies within few cm of propagation.

In the simulation, the parameters of the drive beam are as follows: $190~\mathrm{pC}$ charge,
$18~\mathrm{fs}$ duration and a peak current of $10~\mathrm{kA}$.
Based on the LWFA simulation presented in section~\ref{sec:lwfa} and supported
by the experiments performed at HZDR with a $98~\mathrm{TW}$ laser pulse~\cite{Couperus2017,Irman2018},
we assume that an electron beam with these characteristics could be produced by a near PW-class laser system
operating an LWFA at a plasma density close to $2 \times 10^{17}~\mathrm{cm^{-3}}$ (see section~\ref{sec:scaling}).
Thus, we further assume that the average energy of this beam could be $3\,\mathrm{GeV}$,
with $10~\%$ relative energy spread, and that the normalized emittance is $15~\mathrm{\upmu m}$.
For simplicity, the relative energy spread is considered time uncorrelated in the simulation.
The charge distribution of the beam is initialized Gaussian in every phase-space dimension
and cylindrically symmetric with respect to the propagation axis.
The transverse rms size of the beam is $\sigma_z = 1.5~\mathrm{\upmu m}$ at waist, when entering into the plasma.
This assumption neglects any drift between the two plasma stages,
and considers that the laser is completely removed/reflected by a plasma-mirror~\cite{Thaury2007,Steinke2016},
while the electron beam goes through with negligible impact on its transverse size and emittance. 
In the simulation, the plasma profile and the dopant section are also idealized with the aim to
explore the potential of this approach. 
The plasma profile starts with a short Gaussian up-ramp from vacuum
to the plateau density at $n_p = 8 \times 10^{18}~\mathrm{cm^{-3}}$.
Then the plasma density continues uniform at $n_p$ for the rest of the simulation.
The first part of the plasma profile is doped at $6\%$ with helium (He),
up to only $50~\mathrm{\upmu m}$ from the start of the plateau.

\begin{table*}[!t]
  \begin{tabular}{lll}
    &  Driver (before PWFA) & Witness (after PWFA) \\
    \hline\hline
    Charge                 & $190~\mathrm{pC}$   & $11~\mathrm{pC}$  \\
    Average energy         & $3~\mathrm{GeV}$    & $6~\mathrm{GeV} $ \\
    Energy spread          & $10\%$              & $3\%$  \\
    Average sliced energy spread   & $10\%$              & $0.2\%$ \\
    Normalized emittance   & $15~\mathrm{\upmu m}$ & $0.16~\mathrm{\upmu m}$ \\
    Duration (fwhm)        & $18~\mathrm{fs}$    & $0.8~\mathrm{fs}$  \\
    Current                & $10~\mathrm{kA}$    & $15~\mathrm{kA}$  \\
    Brightness             & $8.8 \times 10^{-2}~\mathrm{kA/\upmu m^2}$ & $1.2 \times 10^{3}~\mathrm{kA/\upmu m^2}$ \\
    \hline\hline
  \end{tabular}
  \caption{Drive and witness beam properties before and after the PWFA stage, respectively.}
  \label{tab:pwfasim}
\end{table*}
Figure~\ref{fig:pwfasim}(a) (top) shows a snapshot of the simulation when the drive beam is traversing
the dopant section.
In this simulation both the hydrogen and the first level of helium are assumed preionized,
while the second ionization level of helium ($\mathrm{He^+}$) is kept non ionized.
This could be achieved by means of a counter-propagating low-intensity laser for selective ionization
(cf.~Figure~\ref{fig:scheme}).
The strong accelerating field at the back of the blowout cavity is responsible for the ionization of
$\mathrm{He^+}$ and the subsequent trapping of a high-quality electron beam~\cite{MartinezdelaOssa2013,MartinezdelaOssa2015}.
Figure~\ref{fig:pwfasim}(a) (middle) shows another snapshot of the simulation
after the dopant section has been passed, at a propagation distance of $z = 2.6~\mathrm{mm}$
after the start of the density plateau.
The witness beam features a peak current of $15~\mathrm{kA}$, $11~\mathrm{pC}$ charge
and $0.8~\mathrm{fs}$ duration (fwhm).
Figure~\ref{fig:pwfasim}(a) (bottom) shows the longitudinal electric field also at $z = 2.6~\mathrm{mm}$.
The accelerating field over the witness at this point is $E_{z}^\mathrm{w} \simeq -475~\mathrm{GV/m}$,
while the maximum decelerating field on the driver is $E_{z}^\mathrm{d} \simeq 170~\mathrm{GV/m}$.
The ratio of these two quantities defines the transformer ratio,
$R = |E_{z}^\mathrm{w}/E_{z}^\mathrm{d}| \simeq 2.8$,
which provides a measure of the maximum energy gain of the witness beam as a function of the initial energy
of the driver $\Updelta\gamma^\mathrm{w} = R~\gamma_0^\mathrm{d}$,
under the assumption that the wakefields remain at this state up to the energy depletion of the driver.
The WII injection method exploits precisely this high transformer ratio ($R$ is considered high for values bigger than 2)
in order to induce ionization and trapping from and into the extreme accelerating fields of the plasma wake,
while avoiding any spurious injection caused by the drive beam.
This allows the initial phase-volume of the trapped electrons to be constrained to a well-defined
phase-range of the wakefields, which in turn, results in the generation of high-quality witness beams~\cite{MartinezdelaOssa2013,MartinezdelaOssa2015}.

Figure~\ref{fig:pwfasim}(b) shows the witness beam phase space after $12~\mathrm{mm}$ of propagation.
The average longitudinal momentum of the beam is $\bar{p}_z = 5.05~\mathrm{GeV/c}$,
while the projected normalized emittance is preserved around $150~\mathrm{nm}$ in both transverse planes.
The total relative energy spread of the witness beam is $\sim 3\%$ in total,
but only $\sim0.5\%$ within the fwhm of the energy peak.
The average sliced relative energy spread is around $0.2\%$.
This means that although the high-current witness beam partially flattens the accelerating fields within its
central part by means of beam loading, the residual energy chirp at the head and tail
contributes significantly to the overall energy spread.
Figure~\ref{fig:pwfasim}(c) shows the evolution of the witness beam parameters
as a function of the propagation distance, featuring an essentially constant energy gain rate, 
derived from the unique wakefield stability of PWFAs in the blowout regime.
We observe that the energy of the witness doubles the initial energy of the driver
after around $14~\mathrm{mm}$ of propagation, yielding an average energy of $6~\mathrm{GeV}$.  
The evolution of the relative energy spread reflects stable beam-loading conditions
over the whole propagation distance shown in Figure~\ref{fig:pwfasim}(c).
Also the projected normalized emittance of the witness exhibits great stability,
as a result of the uniformity of the focusing fields and the low energy spread of the beam.
In comparison with the evolution of the witness beam in the LWFA stage (Figure~\ref{fig:lwfasim}(c)), 
the PWFA stage shows a greatly improved stability on the accelerating conditions.


In summary, this simulation result strongly supports the concept that high-current electron beams
produced in LWFAs can indeed drive strong plasma wakefields themselves,
where a new witness of dramatically improved quality can be injected and accelerated to much higher energies.
For the parameters here considered, Table~\ref{tab:pwfasim} summarizes the properties
of the LWFA-produced drive beam against the newly PWFA-produced witness beam.
In particular, the brightness of the witness beam, $B_\mathrm{w}=2I_{\mathrm{w}}/\epsilon_{n,\mathrm{w}}^2$,
is increased by about five orders of magnitude. 

\section{Simulation for the proof-of-concept experiment at HZDR}
Motivated by the promising simulations results shown in section~\ref{sec:concept},
we are currently exploring the experimental feasibility of this concept.
With this purpose, a proof-of-concept experiment has been implemented at HZDR~\cite{Heinemann2017},
using the DRACO~\cite{Schramm2017} laser system for the LWFA stage
and the thereby produced electron beam as driver for a subsequent PWFA stage.
As a proof-of-concept, the first goal of the experiment is to demonstrate
the injection and acceleration of a new witness beam in the PWFA stage driven by the LWFA beam.
Ultimately, the resulting PWFA beam is ought to feature a substantially higher energy and brightness
than the initial LWFA beam. 

\begin{figure}[!t]
\centering\includegraphics[width=0.9\columnwidth]{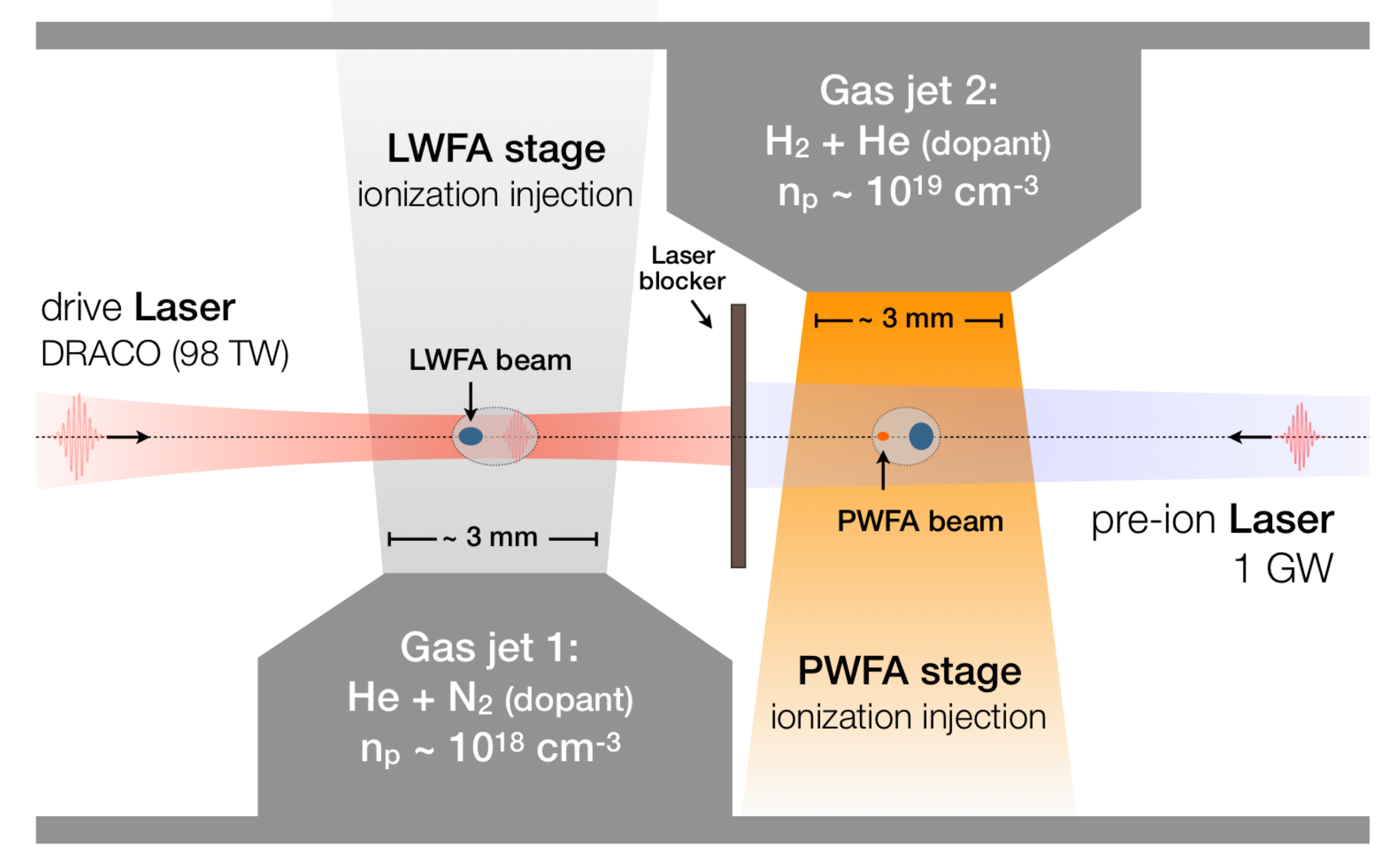}
\caption{Schematic of the double-jet plasma target for the LPWFA proof-of-concept experiment at HZDR.
  In the first gas jet an LWFA stage is driven by the DRACO laser for the generation of a high-current electron beam.
  In the second gas jet, the LWFA-produced electron beam drives a PWFA for the production of a new electron beam
  with largely improved energy and brightness.
  A thin foil made of kapton is placed at the entrance of the second jet in order to reflect the main laser,
  while letting the electron beam go through into the second stage.
  A counter-propagating low-intensity laser can be used in order to fully preionize the hydrogen to facilitate
  the beam refocusing and enhance the blowout formation in the second stage.
}
\label{fig:lpwfahzdr}
\end{figure}

The experimental setup consists of two consecutive supersonic gas jets,
one for the generation of a high-current electron beam in an LWFA stage driven by the DRACO laser,
and a second jet for the injection of a new electron beam in a PWFA stage,
driven by the previously produced electron beam (Figure~\ref{fig:lpwfahzdr}).
A thin ribbon made of kapton, of $15~\mathrm{\upmu m}$ thickness, is placed at the entrance
of the second jet, aiming to reflect the main laser from the second stage,
while letting the electron beam go through with a minimal impact on its transverse size and emittance.
The first LWFA stage has been already proven to provide electron beams of $300~\mathrm{pC}$ charge,
$15\%$ energy spread and with average energies around $250~\mathrm{MeV}$,
with an excellent shot-to-shot stability~\cite{Couperus2017}. 
The second stage will use this LWFA beam to subsequently drive a PWFA,
where a new witness electron beam is generated.

Recent measurements based on coherent transition radiation indicate
that the electron beams produced in the LWFA stage have a duration between 10 and 20~fs,
suggesting peak currents between 15 and 30~kA.
These high currents are more than sufficient to enable a strong blowout wakefield in the second stage,
that allows trapping of electrons generated by field ionization~\cite{MartinezdelaOssa2015}. 
Due to the short duration of the electron beam from the LWFA stage,
the plasma density needs to be increased in the second jet for
near-resonance wakefield excitation, within the $10^{18} - 10^{19}\,\mathrm{cm^{-3}}$ range.
Operating the PWFA near the resonant wakefield excitation 
facilitates the trapping capabilities and maximizes the accelerating gradient~\cite{MartinezdelaOssa2017}. 
One crucial aspect towards the realization of such a staging experiment
is the transition from the LWFA to the PWFA process, particularly the recapturing of
the initially diverging LWFA output in the following plasma section.
This can be achieved by means of the self-driven transverse wakefields in the second plasma.
Assuming that the electron beams leave the first stage at waist,
their transverse rms size, $\sigma_x$, after a drift in vacuum of length $L_\mathrm{drift}$
can be calculated through the expression $\sigma_x^2 = \sigma_{x,0}^2+(L_\mathrm{drift}\,\sigma_{x',0})^2$,
where $\sigma_{x,0}$ and $\sigma_{x',0}$ are the transverse rms size and divergence of the beam
when leaving the plasma. 
After the first jet, the electron beams have been measured to have $\sim7~\mathrm{mrad}$ divergence,
while their transverse size before leaving the plasma is inferred from PIC simulations
and betatron source size measurements~\cite{Kohler2016} to be around $1~\mathrm{\upmu m}$.
This combined determination of the divergence and the spot size of the beam at the focus allows 
an estimation of the normalized emittance of the beam, which yields $\sim 5~\mathrm{\upmu m}$.
Preliminary PIC simulations considering these beams have shown that a separation between jets up
to $1~\mathrm{mm}$ could be tolerated~\cite{Heinemann2017}.
For distances smaller than this, the electron beams emerging from the first stage
are expected to keep its transverse size below $7~\mathrm{\upmu m}$ (rms),
which is still sufficiently small to enable beam self-ionization of the hydrogen gas,
for the generation of the plasma and the self-driven focusing wakefields.
The maximum transverse electric field generated by a beam with a Gaussian and symmetric
transverse distribution is given by
$E_x^\mathrm{max} \simeq 27 (\mathrm{GV/m})\times I_b[\mathrm{kA}]/\sigma_x[\mathrm{\upmu m}]$,
which for a beam with $I_b = 30~\mathrm{kA}$ and $\sigma_x = 7~\mathrm{\upmu m}$ 
yields $\sim 110~\mathrm{GV/m}$, largely above the ionization threshold of hydrogen. 
According to the ADK field-ionization model~\cite{Ammosov1986} adopted in the PIC simulations,
an electric field value of $34~\mathrm{GV/m}$ induces a ionization probability rate of $0.1~\mathrm{fs^{-1}}$ on hydrogen.
Still, a counter-propagating low-intensity laser can be used in order to fully
preionize the hydrogen to facilitate the beam refocusing and enhance the blowout formation
in the second stage.
In addition, the pre-ionization laser can be used for selective ionization
of certain low-ionization threshold levels of the selected dopant species, e.g. the first level of helium.

\begin{figure*}[!t]
\centering\includegraphics[width=1.0\textwidth]{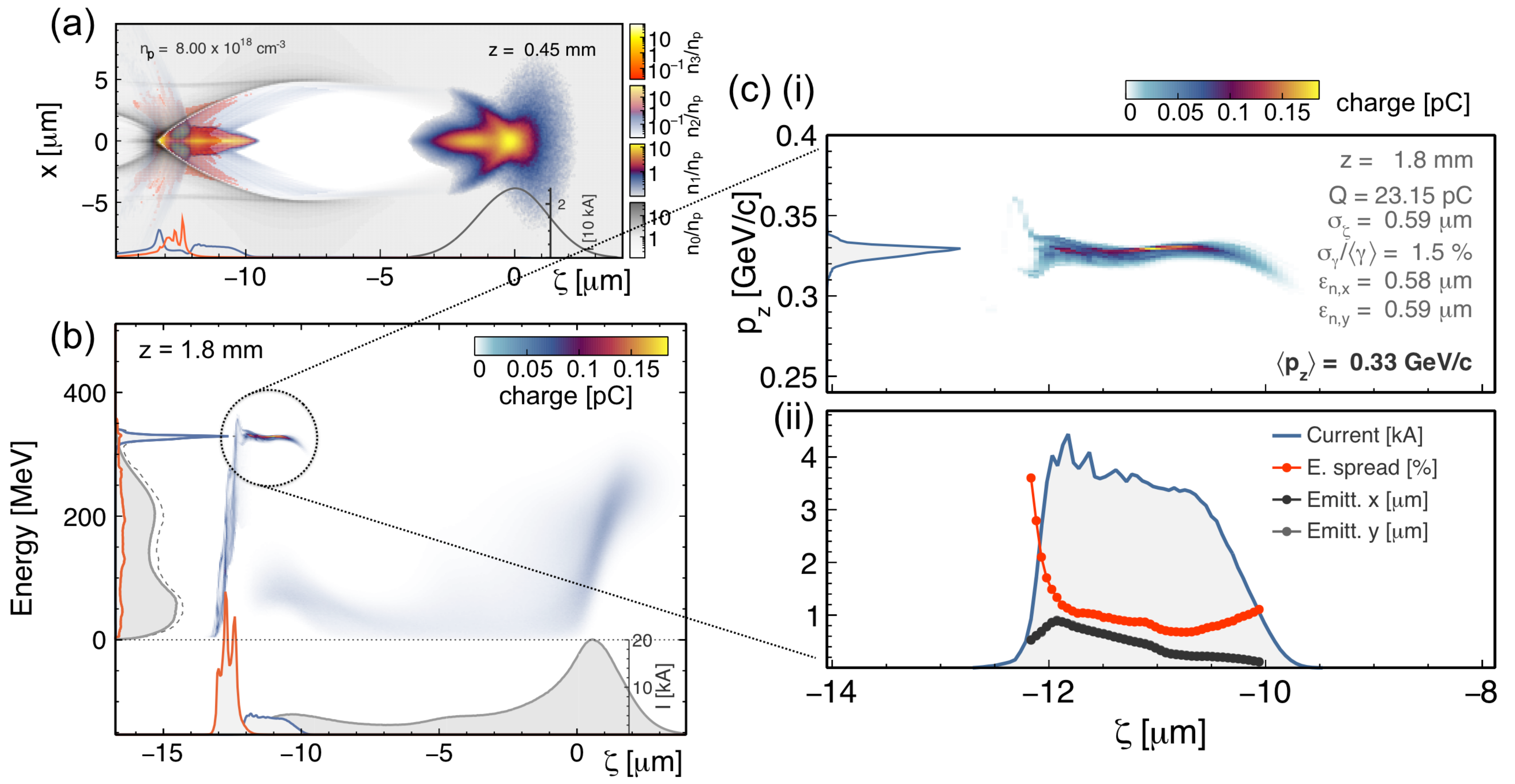}
\caption{3D OSIRIS PIC simulation for the LPWFA stage in the proof-of-concept experiment at HZDR.
  (a): electron density on the central $x-z$ plane of the simulation at $z=0.45~\mathrm{mm}$, 
  for the plasma (gray), the first (blue/yellow) and the second (orange/yellow) electronic levels of helium.
  The dark gray, blue and orange lines at the bottom show the charge per unit of length of the driver,
  electrons from the first and second level of helium, respectively.
  (b): longitudinal phase space of the driver and witness beams after $1.8~\mathrm{mm}$ of propagation.
  (c)~(top): longitudinal phase space of the witness beam after $1.8~\mathrm{mm}$ of propagation,
  and (c)~(bottom): sliced values of the current (blue), relative energy spread (red),
  and normalized emittance in the $x$ (dark grey) and $y$ (light grey) planes.
}
\label{fig:lpwfahzdrsim}
\end{figure*}
In the following we present results from a PIC simulation for the LPWFA stage,
considering the experimental setup described above,
and where the witness electron beam is generated by means of WII injection~\cite{MartinezdelaOssa2013,MartinezdelaOssa2015}.
We note that such an experimental setup could be also compatible with other injection techniques
for the generation of high-quality witness beams~\cite{Hidding2012,Li2013,MartinezdelaOssa2013,MartinezdelaOssa2015,Wittig2015,MartinezdelaOssa2017,Manahan2017}.
In the simulation, the distribution of the beam is Gaussian in every phase-space dimension,
and cylindrically symmetric with respect to the propagation axis. 
The beam parameters are as follows: $252~\mathrm{MeV}$ average energy with $14\%$ energy spread,
$300~\mathrm{pC}$ charge, $10~\mathrm{fs}$ duration, $30~\mathrm{kA}$ peak current and $5~\mathrm{\upmu m}$
normalized emittance (these values are in good agreement with measurements and PIC simulations~\cite{Couperus2017}).
Based on the geometrical constrains of the experiment, we considered a distance between
the two plasma stages of $\sim 700~\mathrm{\upmu m}$, and therefore,
the drive beam is initialized with a transverse size of $5~\mathrm{\upmu m}$,
right at the entrance of the second plasma.
For the plasma target we have considered a longitudinal flat-top profile for the gas distribution
in the second jet.
The gas itself consists of hydrogen doped with helium at $1\%$ concentration.
The hydrogen is considered fully preionized over the total section of the jet
by a large spot size and low intensity laser, capable of fully ionizing the hydrogen
($I_0>1.5\times10^{14}~\mathrm{W/cm^2}$), but not the helium ($I_0<1.14\times10^{15}~\mathrm{W/cm^2}$).
This configuration is different from the simulated case shown in section~\ref{sec:pwfa},
where also the first level of helium was assumed preionized.
The hydrogen plasma density at the plateau is $n_p=8\times10^{18}\mathrm{cm^{-3}}$.
Figure~\ref{fig:lpwfahzdrsim}~(a) shows a snapshot of the simulation after $450~\mathrm{\upmu m}$
of propagation. At this point the drive beam has been fully refocused into the blowout regime.
As a result of this process, an electron bunch composed by electrons ionized from the first
level of helium has been trapped. The trapping of this electron species happens
during the transverse focusing of the beam, which induces a rapid elongation of the blowout cavity.
Due to this cavity elongation, the trapped electrons span a wider phase range, making possible the generation
of longer beams. Another consequence of this process is the self-truncation of the injection. 
In this setup the length of the dopant section is not constrained as in the example shown in section~\ref{sec:pwfa}.
Therefore, it is expected to have continuous injection at the back of the cavity until the accelerating fields
are fully saturated.
In fact, this effect is observed in the simulation for the electrons ionized from the second level of helium.
However, due to the cavity elongation during the focusing,
these electrons are being trapped behind the previously injected beam, and thus,
they do not overlap time nor influence the quality of the primarily injected beam.

Figure~\ref{fig:lpwfahzdrsim}~(b) shows the combined longitudinal phase space of the driver and the
injected electrons, after $1.8~\mathrm{mm}$ of propagation.
At this point the driver is already partly depleted, as a substantial fraction of its electrons
lost their kinetic energy, while the primary injected bunch reaches an average energy of $330~\mathrm{MeV}$
($32\%$ higher than the initial energy of the driver),
within a quasi-monochromatic peak of only $1.6\%$ energy spread.
With more details, we show in Figure~\ref{fig:lpwfahzdrsim}~(c) the longitudinal phase space of the witness beam,
featuring $23~\mathrm{pC}$, $6~\mathrm{fs}$ duration and an approximately flat-top $4~\mathrm{kA}$
current profile. The projected normalized emittance of the beam is $0.5~\mathrm{\upmu m}$,
an order of magnitude smaller than the initial emittance of the driver beam produced in the LWFA stage.
The sliced normalized emittance is also shown in Figure~\ref{fig:lpwfahzdrsim}~(c)~(bottom),
exhibiting values of around $100~\mathrm{nm}$ in its frontal part.
The sliced relative energy spread reaches sub-percent levels along the full length of the bunch.

\begin{figure}[!t]
\centering\includegraphics[width=1.0\columnwidth]{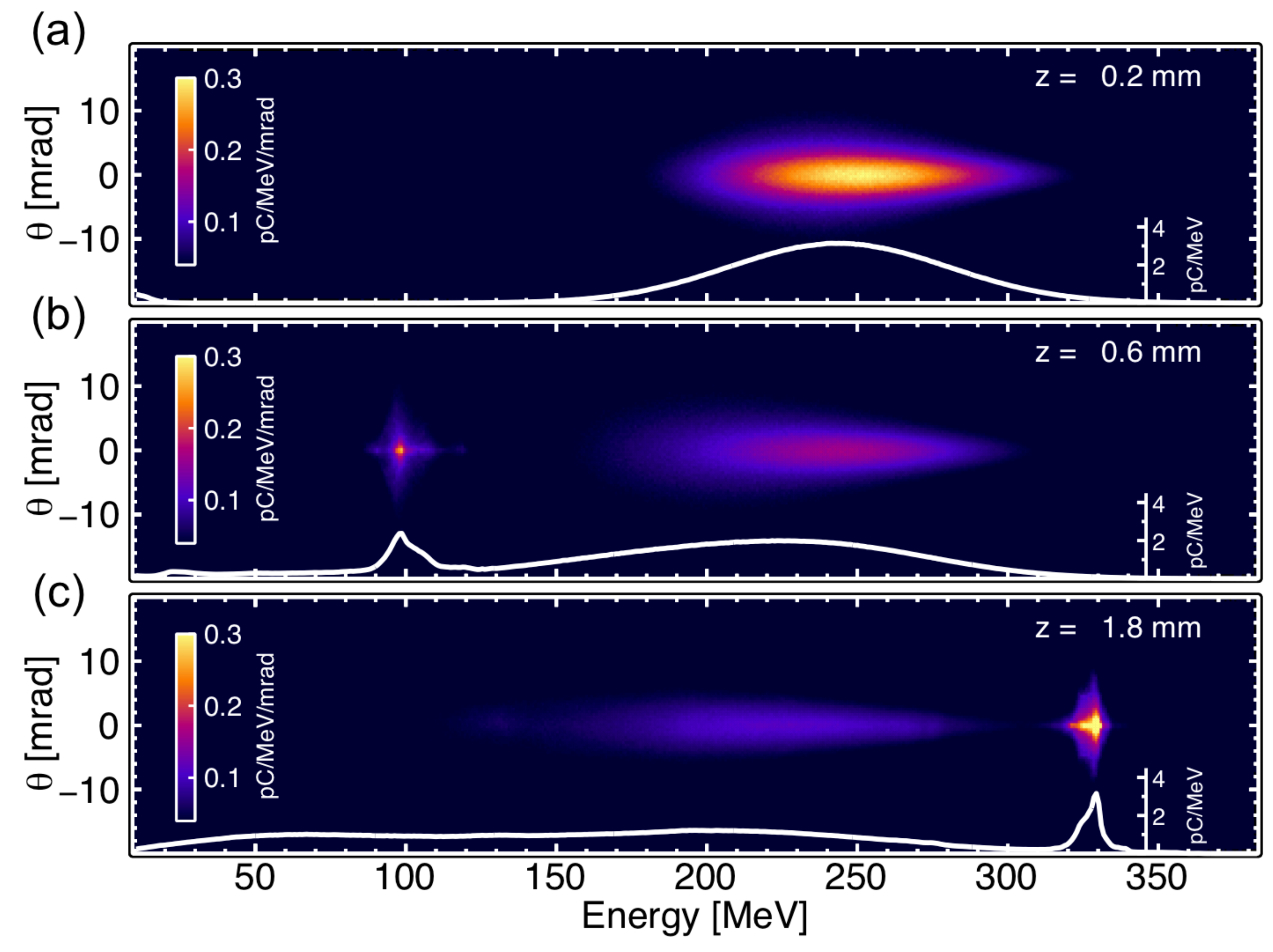}
\caption{3D OSIRIS PIC simulation for the LPWFA stage in the proof-of-concept experiment at HZDR.
  Distribution of the driver and witness beams on the divergence vs. energy plane,
  for three propagation distances. The white curves represent the projections on the energy axis.}
\label{fig:lpwfaspec}
\end{figure}
Figure~\ref{fig:lpwfaspec} shows the distribution of the driver and witness beams
in the divergence vs. energy plane, for three lengths of propagation in the second stage.
Figure~\ref{fig:lpwfaspec}~(a) shows the system after only $200~\mathrm{\upmu m}$ of propagation, 
where we can see the imprint of the large energy spread driver only, peaking at $250~\mathrm{MeV}$ energy.
In Figure~\ref{fig:lpwfaspec}~(b), the newly injected witness beam can be clearly seen
at $\sim100~\mathrm{MeV}$ after $0.6~\mathrm{mm}$ of propagation.
Finally, Figure~\ref{fig:lpwfaspec}~(c) shows 
the witness beam at $330~\mathrm{MeV}$ after $\sim1.8~\mathrm{mm}$
of acceleration, together with the energy depleted drive beam.
The non-overlapping energy distributions of driver and witness allow for
clear observation in experimental realizations.

Experimental studies of the idealized schemes outlined here are of high importance beyond proving
the general feasibility of LPWFAs. In particular, evaluating the sensitivity of the final beam
parameters on the experimental conditions will be critical for assessing the importance of
such schemes for future applications~\cite{Walker2017}.
For this, realistic start-to-end simulations and numerical sensitivity studies
- that are beyond the scope of this work -
will be necessary to evaluate the robustness and scalability of the schemes sketched here.

\section{Conclusion}
In conclusion, we have presented and discussed a conceptual design for an LWFA-driven PWFA (LPWFA),
aiming at producing a new generation of beams with substantially improved energy, and especially, brightness.
The concept is supported by 3D PIC simulations for both LWFA and PWFA stages,
demonstrating that ultra-short and low-emittance beams could be produced in a PWFA stage driven
by a relativistic high-current beam produced in an LWFA.
The expected increase in brightness can be of about five orders of magnitude,
therefore offering an attractive path to the production of GeV-class, ultra-high-brightness beams
for applications in high-energy physics and photon science.
Remarkably, the experimental design is comparably simple, as the two plasma stages can be concatenated
without requiring external coupling elements, thus avoiding complications which would arise from
the implementation of sophisticated beam transport optics between stages.
Finally, we have presented first simulations for the LPWFA proof-of-concept experiment at HZDR,
in which we expect to demonstrate 
that LWFA witness beams can indeed be used as PWFA drivers for the production
of a new class of electron beams with dramatically improved brightness and energy spread.

\vspace{0.5cm}

\begin{description}
\item [Data Accessibility]
  The simulation data supporting this study takes about 1TB of information and
  it is stored in the DESY data servers. 
  It could be accessed through the corresponding author upon reasonable request.
\item [Authors’ Contributions]
  A.M.O. conceived the idea, performed the simulations, the analysis and wrote the manuscript.
  A.F.P. and T.H. contributed to the writing of the manuscript.
  A.D., R.P., A.F.P. and T.H. contributed to the simulation work.
  J.P.C.C., T.H., O.K., T.K., A.M.O. and A.I. designed and developed the LPWFA proof-of-concept experiment at HZDR.
  R.W.A, M.B., S.C., A.D., A.D., M.F.G., B.H., S.K., O.K., J.O. and U.S. reviewed the article and contributed to the final version of the text. 
  All authors gave final approval for publication.
\item [Competing Interests]
  The authors declare no competing interests.
\item [Funding]
  This work used computational time, granted by the
  J\"{u}lich Supercomputing Center, on JUQUEEN and JUWELS machines
  under Projects No. HHH23 and No. HHH45, respectively.
\item [Acknowledgements]
  We thank the OSIRIS consortium (IST/UCLA) for access to the OSIRIS code. 
  Furthermore, we acknowledge the use of the High-Performance Cluster (Maxwell)
  at DESY.
  A.M.O.~acknowledges the Helmholtz Virtual Institute VH-VI-503,
  for financial support, and EuPRAXIA for scientific advisory.
\end{description}

%


\begin{thebibliography}{9}

\bibitem{Tajima1979}
  T. Tajima and J. Dawson,
  \href{https://journals.aps.org/prl/abstract/10.1103/PhysRevLett.43.267}
  {Phys. Rev. Lett. {\bf 43}, 267 (1979)}.
  
\bibitem{Veksler1956}
  V. I. Veksler, \href{http://cds.cern.ch/record/1241563}
  {Conf. Proc. CERN (1956)}.

\bibitem{Chen1985}
  P. Chen, J. M. Dawson, R. W. Huff, and T. Katsouleas,
  \href{https://journals.aps.org/prl/abstract/10.1103/PhysRevLett.54.693}
  {Phys. Rev. Lett. {\bf 54}, 693 (1985)}.

\bibitem{Chen1987}
  P. Chen,
  \href{http://inspirehep.net/record/227160}
  {Part. Accel. {\bf 20}, 171 (1987)}.

\bibitem{Rosenzweig1991}
  J. B. Rosenzweig, B. Breizman, T. C. Katsouleas, and J. J. Su,
  \href{https://journals.aps.org/pra/abstract/10.1103/PhysRevA.44.R6189}
  {Phys. Rev. {\bf A44}, R6189 (1991)}.

\bibitem{Lotov2004}
  K. V. Lotov, 
  \href{https://journals.aps.org/pre/abstract/10.1103/PhysRevE.69.046405}
  {Phys. Rev. {\bf E69}, 046405 (2004)}.

\bibitem{Lu2006}
  W. Lu, C. Huang, M. Zhou, W. B. Mori, and T. Katsouleas,
  \href{https://journals.aps.org/prl/abstract/10.1103/PhysRevLett.96.165002}
  {Phys. Rev. Lett. {\bf 96}, 165002 (2006)}.
  
\bibitem{Thomas2007}
  A. G. R. Thomas, Z. Najmudin, S. P. D. Mangles, C. D. Murphy, A. E. Dangor,
  C. Kamperidis, K. L. Lancaster, W. B. Mori, P. A. Norreys, W. Rozmus, and K. Krushelnick,
  \href{https://journals.aps.org/prl/abstract/10.1103/PhysRevLett.98.095004}
  {Phys. Rev. Lett. {\bf 98}, 095004 (2007)}.

\bibitem{Lu2007}
  W. Lu, M. Tzoufras, and C. Joshi, F. S. Tsung, W. B. Mori, J. Vieira, R. A. Fonseca, and L. O. Silva,
  \href{https://journals.aps.org/prab/abstract/10.1103/PhysRevSTAB.10.061301}
  {Phys. Rev. ST Accel. Beams 10, 061301 (2007)}.

\bibitem{Spence2000}
  D. J. Spence and S. M. Hooker,
  \href{https://doi.org/10.1103/PhysRevE.63.015401}
  {Phys. Rev. E {\bf 63}, 015401 (2000)}.

\bibitem{Shalloo2018}
  R. J. Shalloo, C. Arran, L. Corner, J. Holloway, J. Jonnerby, R. Walczak, H. M. Milchberg, and S. M. Hooker,
  \href{https://journals.aps.org/pre/abstract/10.1103/PhysRevE.97.053203}
  {Phys. Rev. E {\bf 97}, 053203 (2018)}.

\bibitem{Streeter2018}
  M. J. V. Streeter, S. Kneip, M. S. Bloom, R. A. Bendoyro, O. Chekhlov, A. E. Dangor, A. D\"{o}pp, C. J. Hooker,
  J. Holloway, J. Jiang, N. C. Lopes, H. Nakamura, P. A. Norreys, C. A. J. Palmer, P. P. Rajeev, J. Schreiber,
  D. R. Symes, M. Wing, S. P. D. Mangles, and Z. Najmudin,
  \href{https://journals.aps.org/prl/abstract/10.1103/PhysRevLett.120.254801}
  {Phys. Rev. Lett. {\bf 120}, 254801 (2018)}.

\bibitem{Lotov2005}
  K. V. Lotov,
  \href{https://aip.scitation.org/doi/10.1063/1.1889444}
       {Phys. Plasmas {\bf 12}, 053105 (2005)}.

\bibitem{Tzoufras2008}
  M. Tzoufras, W. Lu, F. S. Tsung, C. Huang, W. B. Mori,
  T. Katsouleas, J. Vieira, R. A. Fonseca, and L. O. Silva,       
  \href{https://journals.aps.org/prl/abstract/10.1103/PhysRevLett.101.145002}
       {Phys. Rev. Lett. {\bf 101}, 145002 (2008)}
     
\bibitem{Hidding2012}
  B. Hidding, G. Pretzler, J. B. Rosenzweig, T. K\"{o}nigstein, D. Schiller,
  and D. L. Bruhwiler,
  \href{https://journals.aps.org/prl/abstract/10.1103/PhysRevLett.108.035001}
  {Phys. Rev. Lett. {\bf 108}, 035001 (2012)}.

\bibitem{Li2013}
  F. Li, J. F. Hua, X. L. Xu, C. J. Zhang, L. X. Yan, Y. C. Du, W. H. Huang,
  H. B. Chen, C. X. Tang, W. Lu, C. Joshi, W. B. Mori, and Y. Q. Gu,
  \href{https://journals.aps.org/prl/abstract/10.1103/PhysRevLett.111.015003}
  {Phys. Rev. Lett. {\bf 111}, 015003 (2013)}.
       
\bibitem{MartinezdelaOssa2013}
  A. Martinez de la Ossa, J. Grebenyuk, T. J. Mehrling,  L. Schaper, and J. Osterhoff,
  \href{https://journals.aps.org/prl/abstract/10.1103/PhysRevLett.111.245003}
  {Phys. Rev. Lett. {\bf 111}, 245003 (2013)}.
       
\bibitem{MartinezdelaOssa2015}
  A. Martinez de la Ossa, T. J. Mehrling, L. Schaper, M. J. V. Streeter, and J. Osterhoff,
  \href{https://aip.scitation.org/doi/full/10.1063/1.4929921}
  {Phys. Plasmas {\bf 22} 093107 (2015)}.

\bibitem{Wittig2015}
  G. Wittig, O. Karger, A. Knetsch, Y. Xi, A. Deng, J. B.  Rosenzweig, D. L. Bruhwiler,
  J. Smith, G. G. Manahan, Z.-M. Sheng, D. A. Jaroszynski, and B. Hidding,
  \href{https://journals.aps.org/prab/abstract/10.1103/PhysRevSTAB.18.081304}
       {Phys. Rev. ST Accel. Beams {\bf 18}, 081304 (2015)}.

\bibitem{MartinezdelaOssa2017}
  A. Martinez de la Ossa, Z. Hu, M. J. V. Streeter, T. J. Mehrling,
  O. Kononenko, B. Sheeran, and J. Osterhoff,
  \href{https://journals.aps.org/prab/abstract/10.1103/PhysRevAccelBeams.20.091301}
  {Phys. Rev. Accel. Beams {\bf 20}, 091301 (2017)}.       

\bibitem{Manahan2017}
  G. G. Manahan, A. F. Habib, P. Scherkl, P. Delinikolas, A. Beaton, A. Knetsch, O. Karger, G. Wittig,
  T. Heinemann, Z. M. Sheng, J. R. Cary, D. L. Bruhwiler, J. B. Rosenzweig, and B. Hidding,
  \href{https://www.nature.com/articles/ncomms15705 }
  {Nature Comm. {\bf 8}, 15705 (2017)}.  
  
\bibitem{Hogan2010}
  M. J. Hogan, T. O. Raubenheimer, A. Seryi, P. Muggli, T. Katsouleas, C. Huang,
  W. Lu, W. An, K. A. Marsh, W. B. Mori, C. E. Clayton, and C. Joshi,
  \href{http://iopscience.iop.org/article/10.1088/1367-2630/12/5/055030}
  {New J. Phys. {\bf 12}, 055030 (2010)}.
       
\bibitem{Aschikhin2016}
  A. Aschikhin, C. Behrens, S. Bohlen, J. Dale, N. Delbos, L. di Lucchio, E. Elsen,
  J.-H. Erbe, M. Felber, B. Foster, L. Goldberg, J. Grebenyuk, J.-N. Gruse, B. Hidding,
  Z. Hu, S. Karstensen, A. Knetsch, O. Kononenko, V. Libov, K. Ludwig, A. Maier,
  A. Martinez de la Ossa, T. Mehrling, C. Palmer, F. Pannek, L. Schaper, H. Schlarb,
  B. Schmidt, S. Schreiber, J.-P. Schwinkendorf, H. Steel, M. Streeter,
  G. Tauscher, V. Wacker, S. Weichert, S. Wunderlich, J. Zemella, and J. Osterhoff,
  \href{https://doi.org/10.1016/j.nima.2015.10.005}
  {Nucl. Instrum. Methods Phys. Res. A {\bf 806}, 175 (2016)}.

\bibitem{Strickland1985}
  D. Strickland, and G. Mourou,
  \href{https://www.sciencedirect.com/science/article/pii/0030401885901208}
       {Optics Comm. {\bf 56}, 3 (1985)}.
       
\bibitem{Mangles2004}
  S. P. D. Mangles, C. D. Murphy, Z. Najmudin, A. G. R. Thomas, J. L. Collier, A. E. Dangor, E. J. Divall,
  P. S. Foster, J. G. Gallacher, C. J. Hooker, D. A. Jaroszynski, A. J. Langley, W. B. Mori, P. A. Norreys,
  F. S. Tsung, R. Viskup, B. R. Walton, and K. Krushelnick,
  \href{https://www.nature.com/articles/nature02939}
  {Nature {\bf 431}, 535 (2004)}.
   
\bibitem{Geddes2004}
  C. G. R. Geddes, Cs. Toth, J. van Tilborg, E. Esarey, C. B. Schroeder, D. Bruhwiler, C. Nieter,
  J. Cary, and W. P. Leemans,
  \href{https://www.nature.com/articles/nature02900}
  {Nature {\bf 431}, 538 (2004)}.
   
\bibitem{Faure2004}
  J. Faure, Y. Glinec, A. Pukhov, S. Kiselev, S. Gordienko, E. Lefebvre, J.-P. Rousseau, F. Burgy, and V. Malka,
  \href{https://www.nature.com/articles/nature02963}
  {Nature {\bf 431}, 541 (2004)}.
  
\bibitem{Leemans2006}
  W. P. Leemans, B. Nagler, A. J. Gonsalves, Cs. T\'{o}th, K. Nakamura, C. G. R. Geddes, E. Esarey, C. B. Schroeder,
  and S. M. Hooker,
  \href{https://www.nature.com/articles/nphys418}
  {Nat. Phys. {\bf 2}, 696 (2006)}.

\bibitem{Wang2013}
  X. Wang, R. Zgadzaj, N. Fazel, Z. Li, S. A. Yi, X. Zhang, W. Henderson, Y.-Y. Chang, R. Korzekwa, H.-E. Tsai,
  C.-H. Pai, H. Quevedo, G. Dyer, E. Gaul, M. Martinez, A. C. Bernstein, T. Borger, M. Spinks, M. Donovan,
  V. Khudik, G. Shvets, T. Ditmire, and M. C. Downer,
  \href{https://www.nature.com/articles/ncomms2988}
  {Nature Comm. {\bf 4}, 1988 (2013)}.
  
\bibitem{Leemans2014}
  W. P. Leemans, A. J. Gonsalves, H.-S. Mao, K. Nakamura, C. Benedetti, C. B. Schroeder, Cs. T\'{o}th,
  J. Daniels, D. E. Mittelberger, S. S. Bulanov, J.-L. Vay, C. G. R. Geddes, and E. Esarey,
  \href{https://journals.aps.org/prl/abstract/10.1103/PhysRevLett.113.245002}
  {Phys. Rev. Lett. {\bf 113}, 245002 (2014)}.
  
\bibitem{Osterhoff2008}
  J. Osterhoff, A. Popp, Zs. Major, B. Marx, T. P. Rowlands-Rees, M. Fuchs, M. Geissler, R. H\"{o}rlein,
  B. Hidding, S. Becker, E. A. Peralta, U. Schramm, F. Gr\"{u}ner, D. Habs, F. Krausz, S. M. Hooker, and S. Karsch,
  \href{https://journals.aps.org/prl/abstract/10.1103/PhysRevLett.101.085002}
  {Phys. Rev. Lett. {\bf 101}, 085002 (2008)}.
   
\bibitem{Hafz2008}
  N. A. M. Hafz, T. M. Jeong, I. W. Choi, S. K. Lee, K. H. Pae, V. V. Kulagin, J. H. Sung, T. J. Yu,
  K.-H. Hong, T. Hosokai, J. R. Cary, D.-K. Ko, and J. Lee,
  \href{https://www.nature.com/articles/nphoton.2008.155}
  {Nat. Photon. {\bf 2}, 571 (2008)}.
  
\bibitem{Faure2006}
  J. Faure, C. Rechatin, A. Norlin, A. Lifschitz, Y. Glinec, and V. Malka
  \href{https://www.nature.com/articles/nature05393}
  {Nature {\bf 444}, 737 (2006)}.

\bibitem{Rowlands-Rees2008}
  T. P. Rowlands-Rees, C. Kamperidis, S. Kneip, A. J. Gonsalves, S. P. D. Mangles, J. G. Gallacher, E. Brunetti,
  T. Ibbotson, C. D. Murphy, P. S. Foster, M. J. V. Streeter, F. Budde, P. A. Norreys, D. A. Jaroszynski,
  K. Krushelnick, Z. Najmudin, and S. M. Hooker,
  \href{https://link.aps.org/doi/10.1103/PhysRevLett.100.105005}
       {Phys. Rev. Lett. {\bf 100}, 105005 (2008)}.
  
\bibitem{Pak2010}
  A. Pak, K. A. Marsh, S. F. Martins, W. Lu, W. B. Mori, and C. Joshi,
  \href{https://journals.aps.org/prl/abstract/10.1103/PhysRevLett.104.025003}
  {Phys. Rev. Lett. {\bf 104}, 025003 (2010)}. 
  
\bibitem{Schmid2010}
  K. Schmid, A. Buck, C. M. S. Sears, J. M. Mikhailova, R. Tautz, D. Herrmann, M. Geissler, F. Krausz, and L. Veisz,
  \href{https://journals.aps.org/prab/abstract/10.1103/PhysRevSTAB.13.091301}
  {Phys. Rev. ST Accel. Beams {\bf 13}, 091301 (2010)}.

\bibitem{Gonsalves2011}
  A. J. Gonsalves, K. Nakamura, C. Lin, D. Panasenko, S. Shiraishi, T. Sokollik, C. Benedetti, C. B. Schroeder,
  C. G. R. Geddes, J. van Tilborg, J. Osterhoff, E. Esarey, C. Toth, and W. P. Leemans,
  \href{https://www.nature.com/articles/nphys2071}
  {Nat. Phys. {\bf 7}, 862 (2011)}.

\bibitem{Buck2013}
  A. Buck, J. Wenz, J. Xu, K. Khrennikov, K. Schmid, M. Heigoldt, J. M. Mikhailova, M. Geissler, B. Shen, F. Krausz, S. Karsch, and L. Veisz,
  \href{https://journals.aps.org/prl/abstract/10.1103/PhysRevLett.110.185006}
  {Phys. Rev. Lett. {\bf 110}, 185006 (2010)}.
       
\bibitem{Mirzaie2015}
  M. Mirzaie, S. Li, M. Zeng, N. A. M. Hafz, M. Chen, G. Y. Li, Q. J. Zhu, H. Liao, T. Sokollik, F. Liu,
  Y. Y. Ma, L.M. Chen, Z. M. Sheng, and J. Zhang,
  \href{http://www.nature.com/articles/srep14659}
  {Scientific Reports {\bf 5}, 14659 (2015)}.

\bibitem{Schlenvoigt2008}
  H.-P. Schlenvoigt, K. Haupt, A. Debus, F. Budde, O. J\"{a}ckel, S. Pfotenhauer, H. Schwoerer, E. Rohwer,
  J. G. Gallacher, E. Brunetti, R. P. Shanks, S. M. Wiggins, and D. A. Jaroszynski, 
  \href{https://www.nature.com/articles/nphys811}
  {Nat. Phys. {\bf 4}, 130-133 (2008)}.
  
\bibitem{Fuchs2009}
  M. Fuchs, R. Weingartner, A. Popp, Z. Major, S. Becker, J. Osterhoff, I. Cortrie, B. Zeitler, R. H\"{o}rlein,
  G. D. Tsakiris, U. Schramm, T. P. Rowlands-Rees, S. M. Hooker, D. Habs, F. Krausz, S. Karsch, and F. Gr\"{u}ner,
  \href{https://www.nature.com/articles/nphys1404}
  {Nat. Phys. {\bf 5}, 826 (2009)}.
  
\bibitem{Kneip2010}
  S. Kneip, C. McGuffey, J. L. Martins, S. F. Martins, C. Bellei, V. Chvykov, F. Dollar,
  R. Fonseca, C. Huntington, G. Kalintchenko, A. Maksimchuk, S. P. D. Mangles, T. Matsuoka, S. R. Nagel,
  C. A. J. Palmer, J. Schreiber, K. Ta Phuoc, A. G. R. Thomas, V. Yanovsky, L. O. Silva, K. Krushelnick
  and Z. Najmudin,
  \href{https://www.nature.com/articles/nphys1789}
  {Nat. Phys. {\bf 6}, 980 (2010)}.

\bibitem{Khrennikov2015}
  K. Khrennikov, J. Wenz, A. Buck, J. Xu, M. Heigoldt, L. Veisz, and S. Karsch,
  \href{https://journals.aps.org/prl/abstract/10.1103/PhysRevLett.114.195003}
  {Phys. Rev. Lett. {\bf 114}, 195003 (2015)}.
       
\bibitem{Dopp2017}
  A. D\"{o}pp, B. Mahieu, A. Lifschitz, C. Thaury, A. Doche, E. Guillaume, G. Grittan, O. Lundh, M. Hansson,
  J. Gautier, M. Kozlova, J. P. Goddet, P. Rousseau, A. Tafzi, V. Malka, A. Rousse, S. Corde, and K. T. Phuoc,
  \href{https://www.nature.com/articles/lsa201786}
  {Light: Science \& Applications {\bf 6}, e17086 (2017)}.
       
\bibitem{Blumenfeld2007}
  I. Blumenfeld, C. E. Clayton, F.-J. Decker, M. J. Hogan, C. Huang, R. Ischebeck, R. Iverson, C. Joshi,
  T. Katsouleas, N. Kirby, W. Lu, K. A. Marsh, W. B. Mori, P. Muggli, E. \"{O}z, R. H. Siemann,
  D. Walz, and M. Zhou,
  \href{https://www.nature.com/articles/nature05538}
  {Nature {\bf 445}, 741 (2007)}.

\bibitem{Oz2007}
  E. \"{O}z, S. Deng, T. Katsouleas, P. Muggli, C. D. Barnes, I. Blumenfeld,
  F. J. Decker, P. Emma, M. J. Hogan, R. Ischebeck, R. H. Iverson, N. Kirby,
  P. Krejcik, C. O'Connell, R. H. Siemann, D. Walz, D. Auerbach, C. E. Clayton,
  C. Huang, D. K. Johnson, C. Joshi, W. Lu, K. A. Marsh, W. B. Mori, and M. Zhou,
  \href{https://journals.aps.org/prl/abstract/10.1103/PhysRevLett.98.084801}
  {Phys. Rev. Lett. {\bf 98}, 084801 (2007)}.

\bibitem{Litos2014}
  M. Litos, E. Adli, W. An, C. I. Clarke, C. E. Clayton, S. Corde, J. P. Delahaye, R. J. England, A. S. Fisher,
  J. Frederico, S. Gessner, S. Z. Green, M. J. Hogan, C. Joshi, W. Lu, K. A. Marsh, W. B. Mori, P. Muggli,
  N. Vafaei-Najafabadi, D. Walz, G. White, Z. Wu, V. Yakimenko, and G. Yocky,
  \href{https://www.nature.com/articles/nature13882}
  {Nature {\bf 515}, 92 (2014)}.

\bibitem{Corde2015}
  S. Corde, E. Adli, J. M. Allen, W. An, C. I. Clarke, C. E. Clayton, J. P. Delahaye, J. Frederico, S. Gessner,
  S. Z. Green, M. J. Hogan, C. Joshi, N. Lipkowitz, M. Litos, W. Lu, K. A. Marsh, W. B. Mori, M. Schmeltz,
  N. Vafaei-Najafabadi, D. Walz, V. Yakimenko, and G. Yocky,
  \href{https://www.nature.com/articles/nature14890}
  {Nature {\bf 524}, 442 (2015)}.
  
\bibitem{Corde2016}
  S. Corde, E. Adli, J. M. Allen, W. An, C. I. Clarke, B. Clausse, C. E. Clayton, J. P. Delahaye, J. Frederico,
  S. Gessner, S. Z. Green, M. J. Hogan, C. Joshi, M. Litos, W. Lu, K. A. Marsh, W. B. Mori, N. Vafaei-Najafabadi,
  D. Walz, and V. Yakimenko,
  \href{https://www.nature.com/articles/ncomms11898}
  {Nature Comm. {\bf 7}, 11898 (2016)}.

\bibitem{Gross2018}
  M. Gross, J. Engel, J. Good, H. Huck, I. Isaev, G. Koss, M. Krasilnikov, O. Lishilin, G. Loisch, Y. Renier,
  T. Rublack, F. Stephan, R. Brinkmann, A. Martinez de la Ossa, J. Osterhoff, D. Malyutin, D. Richter,
  T. J. Mehrling, M. Khojoyan, C. B. Schroeder, and F. Gr\"{u}ner,
  \href{https://journals.aps.org/prl/abstract/10.1103/PhysRevLett.120.144802}
  {Phys. Rev. Lett. {\bf 120}, 144802}.

\bibitem{Loisch2018}
  G. Loisch, G. Asova, P. Boonpornprasert, R. Brinkmann, Y. Chen, J. Engel, J. Good, M. Gross, F. Gr\"{u}ner,
  H. Huck, D. Kalantaryan, M. Krasilnikov, O. Lishilin, A. Martinez de la Ossa, T. J. Mehrling, D. Melkumyan,
  A. Oppelt, J. Osterhoff, H. Qian, Y. Renier, F. Stephan, C. Tenholt, V. Wohlfarth, and Q. Zhao,
  \href{https://journals.aps.org/prl/abstract/10.1103/PhysRevLett.121.064801}
  {Phys. Rev. Lett. {\bf 121}, 064801}

\bibitem{Adli2018}
  E. Adli {\sl et al.},
  \href{http://www.nature.com/articles/s41586-018-0485-4}
  {Nature {\bf 561}, 363 (2018)}.

\bibitem{Lundh2011}
  O. Lundh, J. Lim, C. Rechatin, L. Ammoura, A. Ben-Isma\"{i}l, X. Davoine, G. Gallot, J-P. Goddet, E. Lefebvre, V. Malka, and J. Faure
  \href{https://www.nature.com/articles/nphys1872}
  {Nat. Phys. {\bf 7}, 219-222 (2011)}.

\bibitem{Heigoldt2015}
  M. Heigoldt, A. Popp, K. Khrennikov, J. Wenz, S. W. Chou, S. Karsch, S. I. Bajlekov, S. M. Hooker, and B. Schmidt,
  \href{https://journals.aps.org/prab/abstract/10.1103/PhysRevSTAB.18.121302}
  {Phys. Rev. ST Accel. Beams {\bf 18}, 121302 (2015)}.
  
\bibitem{Couperus2017}
  J. P. Couperus, R. Pausch, A. K\"{o}hler, O. Zarini, J.M. Kr\"{a}mer, M. Garten, A. Huebl,
  R. Gebhardt, U. Helbig, S. Bock, K. Zeil, A. Debus, M. Bussmann, U. Schramm, and A. Irman,
  \href{https://www.nature.com/articles/s41467-017-00592-7}
  {Nature Comm. {\bf 8}, 487 (2017)}.

\bibitem{Irman2018}
  A. Irman, J. P. Couperus, A. Debus, A. K\"{o}hler, J. M. Kr\"{a}mer, R. Pausch, O. Zarini, and U. Schramm,
  \href{http://iopscience.iop.org/article/10.1088/1361-6587/aaaef1/meta}
  {Plasma Phys. Control. Fusion {\bf 60} 044015 (2018)}.
  
\bibitem{Li2017}
  Y. F. Li, D. Z. Li, K. Huang, M. Z. Tao, M. H. Li, J. R. Zhao, Y. Ma, X. Guo, J. G. Wang, M. Chen,
  N. Hafz, J. Zhang, and L. M. Chen,
  \href{https://aip.scitation.org/doi/10.1063/1.4975613}
  {Phys. Plasmas {\bf 24}, 023108 (2017)}.
  
\bibitem{Mehrling2017}
  T. J. Mehrling, R. A. Fonseca, A. Martinez de la Ossa, and J. Vieira,
  \href{https://journals.aps.org/prl/abstract/10.1103/PhysRevLett.118.174801}
  {Phys. Rev. Lett. {\bf 118}, 174801 (2017)}.

\bibitem{MartinezdelaOssa2018}
  A. Martinez de la Ossa, T. J. Mehrling, and J. Osterhoff,
  \href{https://journals.aps.org/prl/abstract/10.1103/PhysRevLett.121.064803}
  {Phys. Rev. Lett. {\bf 121}, 064803 (2018)}.
  
\bibitem{Hidding2010}
  B. Hidding, T. K\"{o}nigstein, J. Osterholz, S. Karsch, O. Willi, and G. Pretzler,
  \href{https://journals.aps.org/prl/abstract/10.1103/PhysRevLett.104.195002}
  {Phys. Rev. Lett. {\bf 104}, 195002 (2010)}.

\bibitem{Pae2010}
  K. H. Pae, I. W. Choi, and J. Lee,
  \href{https://aip.scitation.org/doi/10.1063/1.3522757}
  {Phys. Plasmas {\bf 17}, 123104 (2010)}.
  
\bibitem{Masson-Laborde2014}
  P. E. Masson-Laborde, M. Z. Mo, A. Ali, S. Fourmaux, P. Lassonde, J. C. Kieffer, W. Rozmus,
  D. Teychenne, and R. Fedosejevs,
  \href{https://aip.scitation.org/doi/10.1063/1.4903851}
  {Phys. Plasmas {\bf 21}, 123113 (2014)}.
         
\bibitem{Corde2011}
  S. Corde, C. Thaury, K. Ta Phuoc, A. Lifschitz, G. Lambert, J. Faure, O. Lundh,
  E. Benveniste, A. Ben-Ismail, L. Arantchuk, A. Marciniak, A. Stordeur, P. Brijesh,
  A. Rousse, A. Specka, and V. Malka,
  \href{https://journals.aps.org/prl/abstract/10.1103/PhysRevLett.107.215004}
  {Phys. Rev. Lett. {\bf 107}, 215004 (2011)}.
  
\bibitem{Dong2018}
  C. F. Dong, T. Z. Zhao, K. Behm, P. G. Cummings, J. Nees, A. Maksimchuk,
  V. Yanovsky, K. Krushelnick, and A. G. R. Thomas,
  \href{https://journals.aps.org/prab/abstract/10.1103/PhysRevAccelBeams.21.041303}
  {Phys. Rev. Accel. Beams {\bf 21}, 041303 (2018)}.
  
\bibitem{Ferri2018}
  J. Ferri, S. Corde, A. D\"{o}pp, A. Lifschitz, A. Doche, C. Thaury, K. Ta Phuoc, B. Mahieu, I.
  A. Andriyash, V. Malka, and X. Davoine,
  \href{https://journals.aps.org/prl/abstract/10.1103/PhysRevLett.120.254802}
  {Phys. Rev. Lett. {\bf 120}, 254802 (2018)}.

\bibitem{Kuschel2016}
  S. Kuschel, D. Hollatz, T. Heinemann, O. Karger, M. B. Schwab, D. Ullmann, A. Knetsch, A. Seidel,
  C. R\"{o}del, M. Yeung, M. Leier, A. Blinne, H. Ding, T. Kurz, D. J. Corvan, A. S\"{a}vert, S. Karsch,
  M. C. Kaluza, B. Hidding, and M. Zepf
  \href{https://journals.aps.org/prab/abstract/10.1103/PhysRevAccelBeams.19.071301}
  {Phys. Rev. Accel. Beams {\bf 19}, 071301 (2016)}.

\bibitem{Chou2016}
  S. Chou, J. Xu, K. Khrennikov, D. E. Cardenas, J. Wenz, M. Heigoldt, L. Hofmann, L. Veisz, and S. Karsch,
  \href{https://journals.aps.org/prl/abstract/10.1103/PhysRevLett.117.144801}
  {Phys. Rev. Lett. {\bf 117}, 144801 (2016)}.

\bibitem{Gilljohann2018}
  M. F. Gilljohann, H. Ding, A. D\"{o}pp, J. Goetzfried, S. Schindler, G. Schilling,
  S. Corde, A. Debus, T. Heinemann, B. Hidding, S. M. Hooker, A. Irman, O. Kononenko,
  T. Kurz, A. Martinez de la Ossa, U. Schramm, and S. Karsch,
  \href{https://arxiv.org/abs/1810.11813}
  {arXiv:1810.11813 (2018)}.
  
\bibitem{Fonseca2002}
  R. Fonseca, L. Silva, F. Tsung, V. Decyk, W. Lu, C. Ren, W. Mori, S. Deng, S. Lee, T. Katsouleas,
  and J. Adam,
  \href{https://doi.org/10.1007/3-540-47789-6_36}
  {Notes Comp. Sci. {\bf 2331}, 342 (2002)}.

\bibitem{Fonseca2008}
  R. A. Fonseca, S. F. Martins, L. O. Silva, J. W. Tonge, F. S. Tsung, and W. B. Mori,
  \href{https://doi.org/10.1088/0741-3335/50/12/124034}
  {Plasma Phys. Controlled Fusion {\bf 50}, 124034 (2008)}.

\bibitem{Fonseca2013}
  R.A. Fonseca, J. Vieira, F. Fiuza, A. Davidson, F. S. Tsung, W. B. Mori, and L. O. Silva,
  \href{https://doi.org/10.1088/0741-3335/55/12/124011}
  {Plasma Phys. Controlled Fusion {\bf 55}, 124011 (2013)}.

\bibitem{Huang2012}
  Z. Huang, Y. Ding, and C. B. Schroeder
  \href{https://journals.aps.org/prl/abstract/10.1103/PhysRevLett.109.204801}
  {Phys. Rev. Lett. {\bf 109}, 204801 (2012)}.

\bibitem{Maier2012}
  A. R. Maier, A. Meseck, S. Reiche, C. B. Schroeder, T. Seggebrock, and F. Gr\"{u}ner,
  \href{https://journals.aps.org/prx/abstract/10.1103/PhysRevX.2.031019}
  {Phys. Rev. X {\bf 2}, 031019 (2012)}

\bibitem{Hidding2014}
  B. Hidding, G. G. Manahan, O. Karger, A. Knetsch, G. Wittig, D. A. Jaroszynski, Z.-M. Sheng,
  Y. Xi, A. Deng, J. B. Rosenzweig, G. Andonian, A. Murokh, G. Pretzler, D. L. Bruhwiler, and J. Smith,
  \href{http://iopscience.iop.org/article/10.1088/0953-4075/47/23/234010}
  {J. Phys. B: At. Mol. Opt. Phys. {\bf 47} 234010 (2014)}.
  
\bibitem{Dimitri2015}
  S. Di Mitri,
  \href{https://www.mdpi.com/2304-6732/2/2/317}
  {Photonics {\bf 2015}, 2, 317 (2015)}.

\bibitem{Schramm2017}
  U. Schramm {\sl et al.},
  \href{http://iopscience.iop.org/article/10.1088/1742-6596/874/1/012028}
  {J. Phys.: Conf. Ser. {\bf 874} 012028 (2017)}.
  
\bibitem{Walker2017}
  P. A. Walker {\sl et al.},
  \href{http://iopscience.iop.org/article/10.1088/1742-6596/874/1/012029}
  {J. Phys.: Conf. Ser. {\bf 874} 012029 (2017)}.

\bibitem{Kirby2009}
  N. Kirby, I. Blumenfeld, C. E. Clayton, F. J. Decker, M. J. Hogan, C. Huang, R. Ischebeck, R. H. Iverson,
  C. Joshi, T. Katsouleas, W. Lu, K. A. Marsh, S. F. Martins, W. B. Mori, P. Muggli, E. Oz and R. H. Siemann,
  \href{https://journals.aps.org/prab/abstract/10.1103/PhysRevSTAB.12.051302}
  {Phys. Rev. ST Accel. Beams {\bf 12}, 051302 (2009)}.

\bibitem{Schroeder2014}
  C. B. Schroeder, J.-L. Vay, E. Esarey, S. S. Bulanov, C. Benedetti, L.-L. Yu, M. Chen, C. G. R. Geddes,
  and W. P. Leemans,
  \href{https://journals.aps.org/prab/abstract/10.1103/PhysRevSTAB.17.101301}
       {Phys. Rev. ST Accel. Beams {\bf 17}, 101301 (2014)}.  
  
\bibitem{Thaury2007}
  C. Thaury, F. Qu\'{e}r\'{e}, J.-P. Geindre, A. Levy, T. Ceccotti, P. Monot, M. Bougeard,
  F. R\'{e}au, P. d'Oliveira, P. Audebert, R. Marjoribanks, and Ph. Martin,
  \href{http://www.nature.com/articles/nphys595}
  {Nat. Phys. {\bf 3}, 424 (2007)}.
  
\bibitem{Steinke2016}
  S. Steinke, J. van Tilborg, C. benedetti, C. G. R. Geddes, C. B. Schroeder, J. Daniels, K. K. Swanson,
  A. J. Gonsalves, K. Nakamura, N. H. Matlis, b. H. Shaw, E. Esarey, and W. P. Leemans,
  \href{https://www.nature.com/articles/nature16525}
  {Nature {\bf 530}, 190 (2007)}.

\bibitem{Kohler2016}
  A. K\"{o}hler, J. P.Couperus, O. Zarini, A.Jochmann, A. Irman, and U.Schramm,
  \href{https://www.sciencedirect.com/science/article/pii/S0168900216001935}
  {Nucl. Instrum. Methods Phys. Res. A {\bf 829}, 265-269 (2016)}.
  
\bibitem{Heinemann2017}
  T. Heinemann, R. W. Assmann, J. P. Couperus, B. Hidding, A. Knetsch, O. Kononenko,
  A. K\"{o}hler, T. Kurz, U. Schramm, O. Zarini, A. Irman, and A. Martinez de la Ossa,
  \href{http://accelconf.web.cern.ch/AccelConf/ipac2017/papers/tupik010.pdf}
  {Proc. IPAC (2017)}.

\bibitem{Ammosov1986}
  M. V. Ammosov, N. B. Delone, and V. Krainov,
  \href{http://www.jetp.ac.ru/cgi-bin/e/index/e/64/6/p1191?a=list}
       {Sov. Phys. JETP {\bf 64}, 1191 (1986)}.
  
  

\end{thebibliography}

\end{document}